\documentclass[12pt]{article}

\usepackage{graphicx}
\usepackage{latexsym}
\usepackage[left=.95in, top=.95in, right=.95in, bottom=.95in]{geometry}
\usepackage{amsmath, amsthm,amssymb}

\usepackage{booktabs}

\newcommand{\openr}{\hbox{${\rm I\kern-.2em R}$}}
\newcommand{\openn}{\hbox{${\rm I\kern-.2em N}$}}

\newcommand\independent{\protect\mathpalette{\protect\independenT}{\perp}}
    \def\independenT#1#2{\mathrel{\setbox0\hbox{$#1#2$}%
    \copy0\kern-\wd0\mkern4mu\box0}}

\usepackage{color}
\usepackage{url}

\usepackage[numbers, sort&compress]{natbib}
\usepackage{placeins}

\usepackage{multirow}

\newcommand*{\LargerCdot}{\raisebox{0ex}{\scalebox{1.2}{.}}}

\usepackage[font={small,it}]{caption}

\usepackage{changes}

\title{A new approach to hierarchical data analysis: Targeted maximum likelihood estimation for the causal effect of a cluster-level exposure}

 \author{Laura B. Balzer$^{*}$, Wenjing Zheng, Mark J. van der Laan,  \\Maya L. Petersen, and the SEARCH Collaboration \vspace{1em}  \\
{\footnotesize $^*$ Corresponding author; lbalzer@umass.edu}
}
 \date{March 12, 2018}
 
\begin{document}
 \maketitle
 
\begin{abstract}

We often seek to estimate the impact of an exposure naturally occurring or randomly assigned at the cluster-level. For example, the literature  on neighborhood determinants  of health continues to grow. Likewise, community randomized trials are  applied to learn about real-world implementation, sustainability, and population effects of interventions with proven individual-level efficacy. In these settings, individual-level outcomes are correlated due to shared cluster-level factors, including the exposure, as well as  social or biological interactions between individuals. To flexibly and efficiently estimate the effect of a cluster-level exposure, we present two  targeted maximum likelihood estimators (TMLEs). The first TMLE is developed under a non-parametric  causal model, which allows for arbitrary interactions  between individuals within a cluster. These interactions include direct transmission of the outcome (i.e. contagion) and influence of one individual's covariates on another's outcome (i.e. covariate interference). The second TMLE is developed under a causal sub-model assuming  the cluster-level  and individual-specific covariates are sufficient to control for confounding. Simulations compare the alternative estimators and illustrate the potential  gains  from  pairing individual-level risk factors and outcomes during estimation, while avoiding unwarranted assumptions. Our results  suggest that estimation under the   sub-model  can result in bias and misleading inference in an observational setting. Incorporating working assumptions during estimation  is more robust than assuming they hold in the underlying causal model. We illustrate our approach with an application to HIV prevention and treatment.

\end{abstract}



\section{Introduction}

In many studies, individuals are grouped into clusters, such as households,  clinics, or communities, and the objective is to learn the impact of an exposure  naturally occurring or randomly assigned at the cluster-level.  
In observational settings, for example, there is a growing body of literature dedicated  to understanding  neighborhood determinants of health.\citep{Kawachi2003, Oakes2004, Sobel2006}
Likewise, cluster (group) randomized trials are increasingly implemented to learn about  large-scale implementation as well as the direct, indirect, and population-level effects of interventions with proven individual-level efficacy.\citep{HayesMoulton2009}  Examples of ongoing cluster randomized trials include the SEARCH study,  testing a community-based strategy for HIV prevention and treatment;\citep{SEARCH} the CBIM study, testing a school-based program to prevent gender violence;\citep{CoachingBoys2016} and the SHINE study,  testing a household-based strategy to reduce Staphylococcus aureus infection.\citep{SHINE2015} 
In both observational and trial settings, individual-level outcomes may be correlated due to shared cluster-level factors, including the exposure, and causal interactions between individuals within clusters. In this paper, we aim to make full use of a hierarchical data structure to flexibly and efficiently estimate the effect of the cluster-based exposure, while avoiding unwarranted causal and statistical assumptions.

There is an extensive literature on the definition and estimation of the impact of cluster-based exposures or interventions.\citep{HayesMoulton2009, Galbraith2010}
Two popular approaches   are random (mixed) effects models and generalized estimating equations (GEE).\citep{LairdWare82, LiangZeger86} For  reviews of these methods, we refer the reader to Gardiner \emph{et al.} and Hubbard \emph{et al.}, among others.\cite{Gardiner2009,Hubbard2010} In these approaches, the causal effect of interest is defined as the coefficient for exposure in the outcome regression. For estimation and inference, these algorithms harness the pairing of individual-level risk factors and outcomes, while accounting for the correlation of outcomes within clusters.  More recently, 
augmented-GEE has been proposed  to increase precision in cluster randomized trials.\citep{Stephens2012, Stephens2014}
A potential short-coming of these approaches is their reliance on parametric regression models to  define and estimate causal effects. In particular,  background knowledge is rarely sufficient to justify the parametric models employed. In observational settings, this  can result in ill-defined causal effects, biased estimates, and misleading inference.\citep{Hubbard2010}
In cluster randomized trials, this approach can  result in efficiency losses.

In this manuscript, we begin by presenting a structural causal model to represent a general hierarchical data generating process.\citep{Pearl1988, Pearl2000, MarkBook} This causal model is  non-parametric and accounts for dependence in individual-level outcomes that may be induced by shared cluster-level factors and by causal interactions between individuals.\citep{Cox1958, Rubin1980, Halloran1991, Halloran1995, Sobel2006, Hudgens2008}
Throughout we assume independence between clusters. 
The causal model can incorporate, but does not require, assumptions  reflecting  the exposure assignment to clusters  (e.g. randomization). 
Through interventions on this causal model, we generate counterfactuals and define the causal effect of interest without relying on parametric models. 
This approach ensures that the causal effect corresponds to the underlying scientific question and is agnostic to  data generating process (e.g. the presence or absence of informative cluster sizes\citep{Seaman2014}).

If the observed data are aggregated to the cluster-level, then estimation of the corresponding statistical parameter can proceed analogously to non-hierarchical data structures. 
For example, we could apply matching algorithms,\citep{Rosenbaum1983, Abadie&Imbens06, Sekhon2011} parametric G-computation,\citep{Robins1986, Taubman2009, Young2011, Snowden2011} inverse probability of  treatment weighting (IPTW) estimators,\citep{Horvitz1952, Robins2000, Hernan2000, Bodnar2004,Hernan2006b, Shen2014} or double robust approaches,\citep{Robins&Rotnitzky92, Robins1994, vanderLaanRobins2003, Bang&Robins05, vanderLaan2006, MarkBook}  such as targeted maximum likelihood estimation (TMLE). 
This aggregated data approach is straightforward and naturally respects the experimental (independent) unit as the cluster. Furthermore, this approach avoids unwarranted assumptions on the distribution of latent  terms or on the dependence structure within a cluster. 
However, this approach ignores the pairing of the individual-level risk factors with individual-level  outcomes. 

As an alternative to approaches based on aggregated data, we develop two targeted maximum likelihood estimators (TMLE) that leverage the hierarchical data structure by preserving the pairing of individual-level covariates and outcomes.\citep{vanderLaan2006, MarkBook} TMLE is  a general framework for the construction of double robust, semi-parametric, efficient, substitution estimators. 
As applied to causal effect estimation in single time point setting, the algorithm begins with an initial estimator of the \emph{outcome regression}: the conditional mean outcome, given the exposure and baseline covariates. TMLE updates this initial estimator by incorporating information in the known or estimated  \emph{propensity score}: the conditional probability of receiving the exposure, given the covariates. These updated estimates are then plugged into the parameter mapping. 
TMLE is a substitution  estimator, which improves  its stability. 
Through its updating procedures, TMLE  satisfies the efficient score equation, while  guaranteeing parameter estimates respect known bounds (contrary to a direct estimating equation approach).
As a result, TMLE is   {double robust}, yielding a consistent  estimate if either the outcome regression or the propensity score is estimated consistently, and efficient, achieving the lowest possible variance if both the outcome regression and propensity score are estimated consistently at reasonable rates.  
Finally, TMLE naturally integrates machine learning, while maintaining the basis for formal statistical inference. 

In this manuscript, we first propose incorporating the pairing of individual-level covariates and outcomes to improve initial estimation of the outcome regression in a cluster-level TMLE (Section 3). 
Then in Section 4, we  consider assumptions commonly made when estimating effects in hierarchical settings. 
 Specifically, we  assume that an individual's outcome
is generated as a common function of the cluster-level covariates, cluster-level exposure and individual-specific covariates, but is not  
directly affected by the covariates of  other individuals within his/her cluster (i.e. no covariate interference \citep{Prague2016}). 
We further assume that the cluster-level and individual-specific covariates are sufficient to control for confounding. 
%
 For the resulting statistical parameter, 
 we present a   second TMLE for this distinct estimation problem. 
 
 We compare the  two TMLEs theoretically (Section 5) and with finite sample simulations (Section 6). They differ in their efficiency and in how they incorporate individual-level data.
In particular, the assumptions in the more restrictive sub-model result in a lower efficiency bound and thus a potentially more precise TMLE than that 
 developed under the larger model. However,  if these  assumptions do not hold, the  TMLE developed under this sub-model may be subject to bias and misleading inference in a observational setting and to inefficiency in a trial setting.  Since these assumptions are often made when estimating the effects of cluster-level exposures,  our findings may  have implications beyond the Targeted Learning framework.

To illustrate the concepts in this paper, we consider a community-based strategy for intensified HIV testing with immediate initiation of antiretroviral therapy (ART) for all HIV-infected individuals. 
The premise of this
 ``Test-and-Treat"  strategy is to 
 improve clinical outcomes among HIV-infected individuals and dramatically reduce their probability to transmission to others.\citep{Cohen2011,UNAIDS909090,WHO2015guidelines, TEMPRANO,START, Cohen2015} 
 Our objective is estimate the impact  of this  strategy as compared to the standard of care on cumulative HIV incidence: the proportion of baseline HIV-uninfected individuals who become HIV-infected by the end of follow-up.  Within a community, individual outcomes are expected to be correlated due  to both shared community-level factors  and causal interactions between individuals.  The desire to capture the direct, indirect, total, and overall  effects of this nature are a common motivation for focusing on evaluation of cluster-level rather than individual-level interventions.\citep{Rubin1980, Halloran1991, Halloran1995,  Sobel2006, Hudgens2008, HayesMoulton2009}

\section{General hierarchical causal model}\label{EstimationGeneral}

We begin by specifying a structural causal model  for the process that generated data on each cluster (the experimental unit).\cite{Pearl1988, Pearl2000} 
Throughout, we focus on the simple scenario where  a cluster  is first sampled from some target population, and then individuals within a cluster are selected for participation. 
In the running example,  a study community is   randomly selected from the target population of communities, and then baseline HIV-uninfected individuals are randomly sampled from that community.
The number of individuals selected in each cluster could be fixed or could vary. The latter case may arise if  underlying cluster sizes differ and all eligible individuals in each cluster are selected. 
Throughout, clusters are indexed with $j=\{1, \ldots, J\}$, and individuals are indexed with $i=\{1, \ldots, N_j\}$.

After selection of the study units,  covariates are measured. 
These {baseline} characteristics may affect, but are not themselves affected by, the exposure. 
 Some characteristics might be aggregates of individual-level covariates, while others may be cluster-level covariates with no clear individual-level counterpart. 
 The baseline characteristics are divided  into two mutually exclusive sets. For  cluster $j$, let $E_j$ denote the vector of   environmental factors shared by all cluster members, and $\textbf{W}_j$  the matrix of  individual-level characteristics.  In our  example, $E_j$ could include baseline HIV prevalence and community size, while individual-level covariates $\mathbf{W}_j$ might include baseline risk behaviors and demographic data, such as age, sex, and marital status. If there are $p$ such individual-level covariates, then $\mathbf{W}_j$ would be an $(N_j \times p)$ matrix and $W_{ij}$ would be the $(1 \times p)$ vector of baseline characteristics for subject $i$ in cluster $j$. Throughout  $W_{i\LargerCdot}$ denotes the $i^{th}$ individual's covariates from a randomly selected (or unspecified) cluster from the target population. 

Next the exposure $A$ is assigned or  naturally occurs in  each cluster. 
 In our  example, $A_j$ is an indicator that the Test-and-Treat strategy is implemented in  community $j$. 
 The exposure received by  cluster $j$ might be randomly assigned or might depend on  the covariates $(E_j, \mathbf{W}_j)$.
Finally, the outcome  $\mathbf{Y}_j=(Y_{ij}: i=1, \ldots, N_j)$ is measured on all selected individuals in cluster $j$. Throughout, $Y_{i\LargerCdot}$  denotes the $i^{th}$ individual's outcome from a randomly selected (or unspecified) cluster.  In the example,  $Y_{i\LargerCdot}$ is an indicator that  individual $i$  becomes HIV-infected by the end of follow-up.

\begin{figure}
\centering
\includegraphics[width=.3\textwidth]{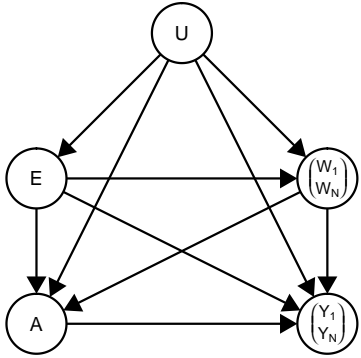}
\caption{Directed acyclic graph for a general hierarchical causal model (Eq.~\ref{SCMcohort}) with $U$ as unmeasured factors, $E$ as cluster-level covariates, $(W_{1.},\ldots W_{N.})$ as individual-level covariates, $A$ as the cluster-level exposure, and $(Y_{1.},\ldots, Y_{N.})$ as individual-level outcomes.
Identifiability (Sec.~\ref{sec.ident}) will require additional assumptions on the unmeasured factors $U$ (Supplementary Figure~S1).}
\label{Figure1}
\end{figure}

Causal relationships between these variables are specified through a directed acyclic graph (Figure~\ref{Figure1}) or non-parametric structural equations:\citep{Pearl1988,Pearl2000}
\begin{align}\label{SCMcohort}
E&=f_E(U_E)\\ \numberwithin{equation}{section}
{\bf W}&=f_{\bf W}(E,U_{\bf W})\nonumber \\ 
A&=f_A(E,{\bf W},U_A) \nonumber \\ 
{\bf Y}&=f_{\bf Y}(E,{\bf W}, A,U_{\bf Y}) \nonumber
\end{align}
where 
$U=(U_E,U_{{\bf W}},U_A,U_{{\bf Y}})$ denotes the  set of 
unmeasured variables.  
This model states that the value of each variable on the left hand side of an equation may be causally determined by the variables on the right hand side of the equation, including  unmeasured sources of random variation $U$. 
Model (\ref{SCMcohort}) contains data generating structures corresponding to both  randomized trials and observational settings. See Appendix A for further details.

This model accounts for many possible sources of dependence between individuals within a cluster. 
For example,  individual-level variables (covariates and  outcomes) will be correlated due to shared measured and unmeasured  factors. The model also allowed for contagion: when an individual's outcome $Y_{ij}$ may be affected by another's outcome $Y_{kj}$ within cluster $j$.\citep{Halloran1991}  Covariate interference  is also consistent with this model: an individual's outcome $Y_{ij}$ may be affected by another's covariates $W_{kj}$.\citep{Prague2016}
No assumptions are made about the structure or correlation of the unmeasured factors $(U_{\mathbf{W}}, U_{\mathbf{Y}})$ between individuals within a cluster.  Thus, this general causal model covers a wide range of
``dependent happenings".\citep{Halloran1991}
It does, however, assume causal independence between distinct clusters (communities).

\subsection{Counterfactuals and the target causal effect}

Counterfactual outcomes are defined through modifications to the data generating process described by   causal model (\ref{SCMcohort}).\citep{Pearl2000, MarkBook} Replacing the structural equation $f_A$ with the constant $a$ generates the counterfactual random variable $\mathbf{Y}(a)$.
Under assumptions linking the structural causal model to the observed data (stated explicitly below), $\mathbf{Y}_j(a)=(Y_{ij}(a):  i=1, \ldots, N_j)$  can be interpreted as the vector of individual-level outcomes that would be observed for cluster $j$ under exposure level $a$. As before, $Y_{i\LargerCdot}(a)$  denotes the $i^{th}$ individual's counterfactual outcome for a randomly selected (unspecified) cluster.
In the  example, $Y_{i\LargerCdot}(1)$ represents the final HIV status for subject $i$ were  his/her community  to receive the Test-and-Treat strategy, irrespective of whether or not the community in fact received the intervention. Likewise, $Y_{i\LargerCdot}(0)$ represents the final HIV  status for subject $i$ were his/her community to continue with the standard of care. 

Let the cluster-level counterfactual outcome be   the (weighted) mean outcome for the $N_j$ individuals sampled from cluster $j$: 
\begin{equation}
 Y^{c}_j(a) \equiv \sum_{i=1}^{N_j} \alpha_{ij} Y_{ij}(a)
 \end{equation}
for some user-specified set of weights such that $\sum_{i=1}^{N_j}\alpha_{ij}=1$. 
When the sample size $N_j$ varies, a natural choice for the weights is the inverse cluster-specific sample size: $\alpha_{ij} = 1/N_j$. 
When the individual-level index $i$ is  informative (e.g.  in a repeated measures setting), other choices of the weight vector $\alpha$ might be preferred. 
 To simplify exposition for the remainder of the article, we assume the weight $\alpha_{ij} = 1/N_j$ and the cluster-level outcome is the empirical mean of the individual-level outcomes.  
   In the running example,  $Y^{c}(a)$ is  the counterfactual proportion of baseline HIV-uninfected individuals who would seroconvert during follow-up if the community received intervention  $A=a$. In other words, $Y^{c}(a)$ is the counterfactual cumulative HIV incidence under exposure level $A=a$.


We focus on causal parameters defined in terms of the \emph{treatment-specific mean}, the expected counterfactual cluster-level outcome if all clusters in the target population received the exposure $A=a$:
$\mathbb{E}[ Y^{c}(a)\big]$.
The difference or ratio of these treatment-specific means defines a causal effect.
For example, the {population average treatment effect} is given by  $\mathbb{E} \big[ Y^{c}(1) \big] -  \mathbb{E}\big[ Y^{c}(0) \big]$.
For the running example, this causal effect evaluates the  difference in  HIV incidence if all communities in our target population implemented the Test-and-Treat strategy versus  if all communities continued with the
standard of care. Alternatively, we could define our parameter of interest as the {causal risk ratio}: $\mathbb{P}(Y^{c}(1)=1)/\mathbb{P}(Y^{c}(0)=1)$.
For conditions and interpretation in terms of a pooled individual-level causal effect, see Appendix B.

\subsection{Observed data and statistical model}

For a randomly sampled cluster, the observed data are the measured environmental covariates, the measured individual-level covariates, the exposure assignment, and the vector of individual-level outcomes: $O = (E, \textbf{W}, A, \textbf{Y})$. 
We define the observed cluster-level outcome as the empirical mean of the individual-level outcomes: 
$Y_j^{c} \equiv \sum_{i=1}^{N_j} \alpha_{ij} Y_{ij}$ 
with our choice of weights $\alpha_{ij}=1/N_j$.
 We assume that the observed data $O_j: j=1, \ldots, J$ are generated by sampling $J$ independent times from some distribution compatible with the causal model.
Thereby,  causal model (\ref{SCMcohort}) implies  a statistical model, which describes the set of possible distributions of $O$ and is denoted  $\mathcal{M}^I$. In many cases, the causal model does not place any restrictions on the set of observed data distributions, and the resulting statistical model is non-parametric. In other cases, such as a randomized trial, knowledge about the exposure assignment mechanism implies a semi-parametric statistical model.
We use subscript 0 to denote the true distributions. The true distribution of the observed data, denoted $\mathbb{P}_0$, 
is an element of the statistical model $\mathcal{M}^I$.

\subsection{Identifiability of the cluster-level causal effect}\label{sec.ident}

 To write the treatment-specific mean as a function of the observed data distribution, we make two additional assumptions, analogous to common identifiability assumptions for non-hierarchical causal effects.\citep{Robins1986}
First, we assume that all the common causes of the cluster-based exposure $A$ and the vector of individual-level outcomes $\mathbf{Y}$ are captured by the measured covariates $(E, \textbf{W})$ (Supplementary Figure~S1). In other words, we assume there is no unmeasured confounding: 
$A \independent \textbf{Y}(a) \mid E,{\bf W}$.
In the HIV example, this assumption would hold by design if the Test-and-Treat intervention were randomly allocated among communities. Otherwise, measuring a rich set of determinants of HIV infection will increase the plausibility of this assumption. 

We also need the positivity assumption (a.k.a. experimental treatment assignment assumption), which ensures that there is sufficient variability in the exposure value within all possible confounder strata: 
$\mathbb{P}_0(A=a \mid E=e, \mathbf{W}= \mathbf{w})> 0 \  a.e.$. 
Under these assumptions, we  have the hierarchical analogue to the  G-computation identifiability result:\citep{Robins1986}
$\mathbb{E} \big[\textbf{Y}(a) \big]= \mathbb{E}_0 \big[ \mathbb{E}_0({\bf Y}\mid A=a,E,{\bf W}) \big]$.
%
This provides us with a general identifiability result for the causal effect of cluster-level exposure $a$ on any cluster-level outcome $Y^{c}$, 
which is some real valued function of the outcome vector ${\bf Y}$:
\begin{eqnarray}
 \Psi^I(\mathbb{P}_0)(a) \equiv \mathbb{E}_0 \big[ \mathbb{E}_0(Y^{c} \mid A=a,E,{\bf W}) \big].
 \label{Eq:Psi0Big}
\end{eqnarray}
We can interpret the resulting statistical estimand as 
 the expected cluster-level outcome, given the exposure and covariates, averaged 
(standardized) with respect to the covariate distribution in the population.  

The randomization and positivity assumptions thus allow us to identify parameters of $\mathbb{E}[Y^c(a)]$, such as the 
 population average treatment effect:
$ \mathbb{E}_0 \big[ \mathbb{E}_0(Y^{c} \mid A=1,E,{\bf W}) - \mathbb{E}_0(Y^{c} \mid A=0,E,{\bf W}) \big]$. 
Likewise, for a binary outcome we can identify the causal risk ratio as $\mathbb{E}_0 \big[ \mathbb{E}_0(Y^{c} \mid A=1,E,{\bf W})\big]/\mathbb{E}_0 \big[ \mathbb{E}_0(Y^{c} \mid A=0,E,{\bf W}) \big]$.

\section{Estimation under the general hierarchical causal model}\label{ClusterLevelTMLE}

In the previous section, we defined the statistical estimand  as a mapping from the statistical model to the parameter space: $\Psi^I:\mathcal{M}^I \rightarrow\openr$.  Under the above randomization and positivity assumptions,
the target parameter $\Psi^I(\mathbb{P}_0)(a)$ 
corresponds to the treatment-specific mean $\mathbb{E}[Y^{c}(a)]$, which can be used to define   both  absolute and relative effects.\citep{Moore2009}

 In this section, we review a targeted maximum likelihood estimator (TMLE) of the statistical parameter $\Psi^I(\mathbb{P}_0)(a)$ (Eq.~\ref{Eq:Psi0Big})  based on 
$J$ i.i.d. observations $O$ from $\mathbb{P}_0 \in \mathcal{M}^I$. 
The efficient influence curve and cluster-level TMLE presented in this section are direct analogs to the standard individual-level TMLE described and implemented elsewhere. \citep{MarkBook, tmlePackage, ltmlePackage}
We then  discuss several approaches for nuisance parameter estimation.
 Our  contribution is to consider candidate estimators making full use of the  hierarchical  data structure (i.e. the pairing of individual-level risk factors and outcomes) during initial estimation of the conditional mean outcome. 

Before proceeding, we introduce some additional notation.
Let us denote the marginal distribution of the baseline covariates as $Q_{E,\mathbf{W}} \equiv \mathbb{P}(E,\mathbf{W})$ and  the conditional mean of the cluster-level outcome given the exposure and covariates as $\bar{Q}^c(A,E,{\bf W})\equiv \mathbb{E}(Y^{c}|A,E,{\bf W})$.
The statistical parameter can thus be written as $\mathbb{E}[\mathbb{E}(Y^c|a,E,\mathbf{W})]=\sum_{e,\mathbf{w}} \bar{Q}^c(a,e,\mathbf{w}) \mathbb{P}(e,\mathbf{w})$, where the summation generalizes to the integral for continuous covariates. 
This clarifies that the statistical parameter only depends on the observed data distribution through $Q = (Q_{E,{\bf W}},\bar{Q}^c)$. 
For the targeting step, we will also need to estimate  the cluster-level propensity score, denoted as $g^c(a | E, \mathbf{W}) \equiv \mathbb{P}(A=a | E, \mathbf{W})$. Without loss of generality, we  assume that the cluster-level outcome $Y^{c}$ is bounded between zero and one.
\citep{Gruber2010} In the running example, $Y^{c}$ is  cumulative HIV incidence and thus a proportion.

\subsection{The cluster-level TMLE}

The efficient influence curve of $\Psi^I$ at $\mathbb{P}_0$ is given by 
\begin{eqnarray}\label{effic_com}
D^{I}(\mathbb{P}_0)(O) &=& \frac{\mathbb{I}(A=a)}{g_0^c(A\mid E,{\bf W})} \big( Y^{c}-\bar{Q}_{0}^c(A,E,{\bf W}) \big) + \bar{Q}_{0}^c(a,E,{\bf W})-\Psi^I(\mathbb{P}_0)(a).
\end{eqnarray}
where $\bar{Q}_0^c(A,E,\mathbf{W})$ denotes the true conditional mean of the cluster-level outcome and $g_0^c(A|E,\mathbf{W})$ denotes the true  cluster-level propensity score. This is a direct analog of the efficient influence curve for the G-computation identifiability result for a non-hierarchical data setting.\citep{vanderLaan2006, MarkBook}
The first component is the weighted deviations between the cluster-level outcome and its expectation given the exposure and covariates; the weights are the inverse of the cluster-level propensity score. The second component  is the deviation between the conditional mean outcome and its expectation over the covariate distribution. 
 The efficient score equation can be generated as  a score of a fluctuation of the covariate distribution and the conditional distribution of the cluster-level outcome, given the exposure and covariates. 
This is used  in formulation of the targeting step in the TMLE.\citep{vanderLaan2006, MarkBook}

 Specifically, suppose we have an initial estimator $\hat{\bar{Q}}^c(A,E,\mathbf{W})$ of the expected cluster-level  outcome $\bar{Q}_0^c(A,E, \mathbf{W})$.  The TMLE algorithm updates this initial estimator with information contained in the known or estimated propensity score $\hat{g}^c(a\mid E,\mathbf{W})$. 
To do so, we 
minimize a pre-specified loss  function along a least favorable (with respect to the statistical estimand) sub-model through $\hat{\bar{Q}}^c(A,E,\mathbf{W})$. 
We choose  the negative log-likelihood loss function
\begin{equation}
-\mathcal{L}^c(\bar{Q}^c)(O)=Y^{c}\log [\bar{Q}^c(A,E,{\bf W})]+(1-Y^{c})\log [1-\bar{Q}^c(A,E,{\bf W})]
\end{equation}
and the logistic sub-model with fluctuation parameter $\epsilon$: 
\begin{equation}\label{fluc_c}
logit[\hat{\bar{Q}}^{c}(\epsilon)]=logit[\hat{\bar{Q}}^{c}]+\epsilon \hat{H}^c
\end{equation}
where $logit(\LargerCdot)=log(\LargerCdot/1-\LargerCdot)$ and 
the ``clever covariate" 
$\hat{H}^c=\frac{\mathbb{I}(A=a)}{\hat{g}^c(a\mid E,{\bf W})}$. 
At zero fluctuation, the initial estimator is returned: $\hat{\bar{Q}}^{c}(\epsilon=0)=\hat{\bar{Q}}^{c}$. 
Furthermore, the score spans the relevant  component of the efficient influence curve (Eq.~\ref{effic_com})   at any distribution $\mathbb{P}$ in our model $\mathcal{M}^I$.

The parametric regression  (Eq.~\ref{fluc_c}) is used to target the initial estimator $\hat{\bar{Q}}^{c}(A,E,\mathbf{W})$ of outcome regression. The amount of fluctuation (i.e. the coefficient $\epsilon$) is estimated with maximum likelihood. Specifically, we run  logistic regression of the cluster-level outcome $Y^{c}$ on the clever covariate $\hat{H}^c$ 
and use  the $logit$ of the initial estimator $\hat{\bar{Q}}^{c}(A,E,\mathbf{W})$ as offset.
Plugging in the estimated fluctuation parameter $\hat{\epsilon}$ provides an updated fit to the outcome regression: 
\begin{equation}
\hat{\bar{Q}}^{c*}=\hat{\bar{Q}}^{c}(\hat{\epsilon}) = logit^{-1}\big[ logit(\hat{\bar{Q}}^{c})+\hat{\epsilon}  \hat{H}^c\big].
\end{equation}
%
%
As  estimator of the covariate distribution, we use the empirical  $\hat{Q}_{E,\mathbf{W}}$, which puts weight $1/J$ on each cluster.
As detailed  in Rose and van der Laan,\cite{Chpt5} the empirical distribution solves the relevant score equation (i.e. relevant component of the efficient influence curve) and does not need  targeting, even in high dimensional settings.\cite{vanderLaan2014bigdata}

   The TMLE 
is  the substitution estimator obtained by plugging $\hat{Q}^*=(\hat{Q}_{E,\mathbf{W}},\hat{\bar{Q}}^{c*})$ into the parameter mapping $\Psi^I$:
\begin{equation}
\Psi^I(\hat{Q}^*)(a)=\frac{1}{J}\sum_{j=1}^J \hat{\bar{Q}}^{c*}(a,E_j,{\bf W}_j).
\end{equation}
The point estimate, denoted $\hat{\psi}^I(a)$,   is the sample average of the targeted predictions of the cluster-level outcome, given the exposure of interest $(A=a)$ and the measured covariates. 

By construction,   TMLE  solves the efficient score equation: $0=\sum_j D^{I}(\hat{Q}^*,\hat{g}^c)$.
As a result, the estimator is double robust in that it remains consistent if only one of the nuisance parameters (the outcome regression or the propensity score) is consistently estimated. 
In an observational setting, this double robustness property improves 
our chances for obtaining a consistent estimate and valid statistical inference.\citep{Gruber2012}
 In a randomized trial, where the propensity score is known, the double robustness property implies that  the TMLE  will remain unbiased regardless of the outcome regression specification and thereby confers wider flexibility in covariate adjustment to increase efficiency.\citep{Rosenblum2010} 
Furthermore, if both nuisance parameters are consistently estimated at reasonable rates,\citep{MarkBook} then the TMLE is asymptotically linear 
with influence curve equal to the efficient influence curve (Eq.~\ref{effic_com})  and asymptotically efficient.\citep{Bickel1997} In other words, this TMLE achieves the lowest possible asymptotic variance among a large class of estimators. 

Under more general conditions,\citep{MarkBook} TMLE is a regular, asymptotically linear estimator, and the Central Limit Theorem can be used to obtain statistical inference.  
Specifically, let 
\begin{equation}
\hat{D}^{I}(\hat{Q}^*, \hat{g}^c)(O_j)  = \frac{\mathbb{I}(A_j=a)}{\hat{g}^c(A_j\mid E_j,{\bf W}_j)} \big( Y_j^{c}-\hat{\bar{Q}}^{c*}(A_j,E_j,{\bf W}_j) \big) +
\hat{\bar{Q}}^{c*}(a,E_j,{\bf W}_j)- \hat{\psi}^I(a)
\end{equation}
be the plug-in estimator of the  influence curve  for observation $O_j$. 
We obtain a variance estimator with the sample variance  of $\hat{D}^{I}(\hat{Q}^*,\hat{g}^c)$ divided by the number of experimental units: $\hat{\sigma}^2 = Var[\hat{D}^I]/J$.
This variance estimator is used to construct Wald-Type 95\%-confidence intervals and carry out hypothesis tests. 
Under additional assumptions, the non-parametric bootstrap provides an alternative to the influence curve-based inference.

\subsection{Data-adaptive estimation of nuisance parameters}

In most applied settings, 
\emph{a priori}-specification of a correct parametric  regression for  the conditional mean outcome $\bar{Q}_0^c(A,E,{\bf W})$ is  impossible. 
We may know and measure the relevant covariates, but specifying the exact functional form is beyond our knowledge. (Recall our causal model often implies a non-parametric or semi-parametric statistical model.)
In a randomized trial, the propensity score is known (e.g. $g_0^c(a \mid E, \mathbf{W}) = 0.5$) and can be consistently estimated with a parametric regression to improve precision.\cite{MarkRobins, Shen2014, Balzer2016DataAdapt} In observational  settings, however, consistent estimation of the propensity score may present similar challenges. An core feature of  TMLE is the use of  machine learning algorithms for  estimation of  both the outcome regression $\bar{Q}_0^c(a, E, \mathbf{W})$ and the propensity score $g_0^c(a | E, \mathbf{W})$. 

We focus on Super Learner,\citep{SuperLearner, Chpt3}   an ensemble  algorithm.\cite{Wolpert92, Breiman96} Super Learner employs  $V$-fold cross-validation to build a convex combination of algorithm-specific predictions to minimize the cross-validated risk, based on a user-specified loss function. The  library of candidate algorithms can include both parametric models and data-adaptive methods (e.g. stepwise regression, support vector machines,\citep{Cortes&Vapnik95} generalized additive models,\citep{Hastie&Tibshirani90} LASSO\citep{Tibshirani96}-  each with multiple tuning parameters). If a correctly specified parametric model is not included in the library, Super Learner  under minimal conditions 
performs asymptotically as well as an ``oracle selector" that uses the  true distribution $\mathbb{P}_0$ to select the optimal convex combination from the library.\citep{SuperLearner, Chpt3} If a correctly specified parametric model is included in the library, Super Learner still achieves an almost parametric rate of convergence. 

Under our statistical model $\mathcal{M}^I$,  Super Learner for   the outcome regression and  propensity score can be implemented using a cluster-level  loss function (Appendix C). 
 Alternatively, to leverage the pairing of individual-level covariates and outcomes and to reduce the dimensionality of the adjustment set,
 we now consider two working assumptions.
 These assumptions  suggest 
alternative approaches to estimating the cluster-level outcome regression $\bar{Q}_0^c(A,E,\mathbf{W})$ and thereby an expanded Super Learner library.

First,
suppose that an individual's outcome is minimally impacted by the covariates of other individuals in his or her cluster:  
$\mathbb{E}_0(Y_{i\LargerCdot}|  A,E,{\bf W})=  \mathbb{E}_0(Y_{i\LargerCdot} | A,E,W_{i\LargerCdot})$. In other words, consider an exclusion restriction that the $i^{th}$ individual's outcome $Y_{i\LargerCdot}$ is only a function of the matrix ${\bf W}$ through his/her own covariates $W_{i\LargerCdot}$.
Second, suppose that this individual-level regression is common in $i$: 
$\mathbb{E}_0(Y_{i\LargerCdot} | A,E,W_{i\LargerCdot})=\bar{Q}_0(A,E,W)$ for some function $\bar{Q}_0$. 
A common function  is natural when $i$  indexes a random permutation $\{1,\ldots,N\}$. 
 Under these working assumptions, we can rewrite the conditional mean of the cluster-level outcome as
\begin{equation} 
\label{workingmodelindiv}
 \bar{Q}_0^c(A,E,{\bf W})= \sum_{i=1}^N \alpha_{i\LargerCdot} \bar{Q}_0(A,E,W_{i\LargerCdot}).
 \end{equation}
 This suggests a natural estimator for $\bar{Q}_0^c$ based on fitting a single regression of the individual-level outcome $Y$ on the exposure and covariates $(A,E, W)$ and then averaging across individuals  within a cluster.  
 In our HIV example, we could estimate  the expected cumulative HIV incidence $\bar{Q}_0^c$ by (i) pooling individuals across clusters, (ii) fitting a individual-level  outcome regression  with weights $\alpha_{ij}$ and with   terms for the cluster-level exposure, the community's baseline HIV prevalence as well as the individual's age and sex; and (iii) averaging the individual-level predictions within clusters. Corresponding data-adaptive approaches are also possible.

These  working assumptions  can be relaxed by  incorporating  knowledge of the dependence structure between individuals within clusters. 
Suppose we are  able to identify or approximate for each individual $i$ the specific set of individuals ${C}_{i\LargerCdot}$ to which  individual $i$ is ``connected". In other words, ${C}_{i\LargerCdot}$ denotes the subset of individuals who influence the $i^{th}$ individual's outcome $Y_{i\LargerCdot}$. Then we  could pose a more general version of the  working model (Eq.~\ref{workingmodelindiv}) by 
including in the $i^{th}$ individual's covariate  vector the covariates of his/her connections $W_{k\LargerCdot}$ for $k \in{C}_{i\LargerCdot}$.
In the HIV example,  an individual's probability of seroconversion might   depend on his/his own sexual behavior as well as the baseline  behavior of the other individuals in his/her sexual network  
$W_{k\LargerCdot}: k \in {C}_{i\LargerCdot}$.%

In summary,  
the utility of the working assumption (Eq.~\ref{workingmodelindiv})  is to generate an expanded set of candidate estimators of the conditional mean of the cluster-level outcome $\bar{Q}_0^c(A,E,\mathbf{W})$ for inclusion in the  Super Learner library. 
Any $(N\times 1$) individual-level covariate vector can alternatively be included in either the covariate matrix $\mathbf{W}$ or as a 
 cluster-level covariate $E$. 
 Therefore, we can include algorithms  that assume $Y_{i\LargerCdot}$ only depends on $W_{i\LargerCdot}$ for investigator-specified subsets of ${\bf W}$. In other words, this working model allows us to consider a variety of dimension reductions for the adjustment set $(E,\mathbf{W})$.  
Super Learner provides a mechanism to choose between and combine candidate individual-level and cluster-level algorithms in response to the data, thereby optimizing estimator performance. 
Step-by-step implementation of the cluster-level TMLE with Super Learner and  corresponding $\texttt{R}$ code is given in Appendix D.

\section{Hierarchical TMLE when causal dependence is restricted}\label{Sect:CausalSubmodel}

The cluster-level TMLE, presented in the previous section, is developed 
under a general hierarchical causal model  that makes no assumptions about the nature or sources of dependence between individuals within a cluster (Eq.~\ref{SCMcohort}; Fig.~\ref{Figure1}).  For identifiability of the impact of the cluster-level exposure, we assume the cluster-level covariates $E$ and whole matrix of covariates $\mathbf{W}$ are sufficient to control for confounding. For initial estimation of the conditional mean of the cluster-level outcome $\bar{Q}_0^c(A, E, \mathbf{W})$, we consider some additional  working assumptions designed to more fully leverage the hierarchical nature of the data (Eq.~\ref{workingmodelindiv}). These assumptions are treated as ``working" assumptions and are not considered to reflect the underlying causal process. If the propensity score $g_0^c(A|E, \mathbf{W}$) is estimated consistently (as will always be true in a randomized trial), then estimating the outcome regression $\bar{Q}_0^c(A, E, \mathbf{W})$ under  these  working assumptions may improve  asymptotic efficiency as well as finite sample bias and variance;  the better the working assumptions approximate the truth, the better the TMLE will perform. 

In this section, we consider an alternative hierarchical causal model, which restricts the causal dependence of individuals within a cluster. Specifically, we assume that an individual's outcome is known not to be affected by the covariates of other individuals in the cluster.  
This more restrictive causal model implies that  the working assumptions (Eq.~\ref{workingmodelindiv})  
hold, thereby changing the statistical model by restricting the set of allowed distributions for outcome regression $\bar{Q}_0^c$. The modified causal model also results in a distinct identifiability result and corresponding estimand.
Specifically, we now need to assume that the cluster-level covariates $E$ and individual $i$-specific covariates $W_{i\LargerCdot}$ are sufficient to control for confounding.
 For the modified statistical estimation problem
, we present the efficient influence curve 
and the corresponding individual-level TMLE.

\subsection{Restricted hierarchical causal model}

We now consider a causal model assuming each individual's outcome $Y_{i\LargerCdot}$ is drawn from a common (in $i$) distribution depending on the cluster-level covariates $E$, each individual's own covariates $W_{i\LargerCdot}$,  the cluster-level exposure $A$,  and  unmeasured factors $U_{Y_{i\LargerCdot}}$, but  not on the measured covariates of all other individuals in that cluster. 
%
 In other words, we assume  no covariate interference.\citep{Prague2016} We further assume that the cluster-level covariates $E$ and individual $i$-specific covariates $W_{i\LargerCdot}$ are sufficient to control for confounding.   This assumption holds by design in a randomized trial, but is a strong assumption on the distribution of unmeasured factors in an observational setting (Supplementary Figure~S2).

\begin{figure}
\centering
\includegraphics[width=.3\textwidth]{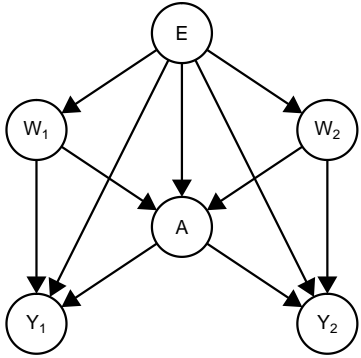}
\caption{Simplified directed acyclic graph for the restricted hierarchical causal model. For ease of presentation, we only show two individuals, denoted by subscripts 1 and 2, in a given cluster and assume all unmeasured factors $U$ are independent. For additional details, see Supplementary Figure~S2.}
\label{Figure2}
\end{figure}

This data generating process is represented by the following structural causal model: 
\begin{align}\label{SCM2}
E&=f_E(U_E) \nonumber \\ 
\mathbf{W} &= f_{\mathbf{W}}(E,U_{\bf W}) \nonumber \\ 
A&=f_A(E, \textbf{W},U_A) \nonumber \\ 
Y_{i\LargerCdot}&=f_{Y}(E,W_{i\LargerCdot}, A, U_{Y_{i\LargerCdot}}), \ i=1,\ldots, N \nonumber \\
\text{where } A & \independent  Y_{i\LargerCdot}(a) \mid E, W_{i\LargerCdot}
\end{align}
We further assume 
that the conditional probability distributions of the individual-level covariates and outcome $(W_{i\LargerCdot}, Y_{i\LargerCdot})$, given  the cluster-level covariates and exposure $(E,A)$, 
  are common in $i$. 
This causal model is compatible with observational studies  (Figure~\ref{Figure2}) and  cluster randomized trials (Supplementary Figure~S3).

Returning to our HIV example,  causal model (\ref{SCM2}) assumes individual $i$'s final HIV status ${Y}_{i\LargerCdot}$ is generated as a common function of the shared environmental factors $E$ (e.g. region, baseline prevalence),  his/her own covariates $W_{i\LargerCdot}$ (e.g. age, sex, marital status), implementation of the Test-and-Treat strategy $A$, and  unmeasured individual-level factors $U_{Y_{i\LargerCdot}}$ (e.g.  his/her perceived stigma), but not the covariates of others in his/her cluster.  
In this infectious disease setting, this causal model might not be realistic. First, the baseline risk behavior  of one individual  $W_{k\LargerCdot}$ may directly or indirectly impact the outcome of another $Y_{i\LargerCdot}$. Even if the assumption of no covariate interference is plausible, this causal model will not hold if there is an unmeasured common cause (e.g. community-level stigma) of the individual-level covariates ($W_{i\LargerCdot}, W_{k\LargerCdot}$) and outcomes ($Y_{i\LargerCdot}, Y_{k\LargerCdot}$).  Of course, we could improve plausibility of these assumptions by including in $W_{i\LargerCdot}$ the baseline covariates of his/her partners $C_{i\LargerCdot}$. Nonetheless, the assumptions in the restricted causal model are commonly made, but potentially implausible when outcomes are biologically or socially transmitted. 
  We refer the reader to Supplementary Figure~S2 for additional examples and discussion.


\subsection{The individual-level TMLE}

Assuming the restricted causal model is true, we proceed to estimation. As before,  the observed data consist of $J$ i.i.d observations of $O=(E, \mathbf{W}, A, \mathbf{Y})$, and the observed cluster-level outcome is the empirical mean of the individual-level outcomes: $Y^{c} = \sum_{i=1}^N \alpha_{i\LargerCdot} Y_{i\LargerCdot}$. 
Without loss of generality, we assume that the individual-level outcome $Y_{i\LargerCdot}$ is bounded in zero and one.
\citep{Gruber2010} In the  example, $Y_{i\LargerCdot}$ is an indicator that the $i^{th}$ individual becomes infected with HIV over the course of follow-up.

Causal model (\ref{SCM2}) implies the statistical assumption in Eq.~\ref{workingmodelindiv}; the conditional mean of the cluster-level outcome can be written as the average of individual-level regressions.
We further assume that the conditional distribution of the exposure, given  the cluster-level  and individual $i$-specific covariates, is a common conditional distribution: 
\begin{equation}
\mathbb{P}_0(A | E, W_{i\LargerCdot}) \equiv g_0(A | E, {W}_{i\LargerCdot}).
\end{equation}
We refer to $g_0(A|E,  {W}_{i\LargerCdot})$ as the individual-level propensity score. 
The resulting statistical model implied by these assumptions is denoted $\mathcal{M}^{II}$ and is a sub-model of $\mathcal{M}^I$.

Under this more restrictive causal model, adjustment  for the cluster-level covariates $E$ and the individual $i$-specific covariates $W_{i\LargerCdot}$ is sufficient to control for  confounding. 
With the corresponding positivity assumption, 
our  identifiability result  for the treatment-specific mean $\mathbb{E}[Y^{c}(a)]$  is given by
\begin{eqnarray}
\Psi^{II}(\mathbb{P}_0)(a) \equiv \mathbb{E}_0\left\{ \sum_{i=1}^N  \alpha_{i\LargerCdot} \bar{Q}_0(a,E,W_{i\LargerCdot})\right\}
\end{eqnarray}
Let 
$\Psi^{II}:\mathcal{M}^{II}\rightarrow\openr$ be the statistical  parameter  implied by this identifiability result, thus defining a new statistical estimation problem. As before, the statistical estimand $\Psi^{II}(\mathbb{P})(a)$ depends on the observed data distribution $\mathbb{P}$ through the marginal distribution of baseline covariates  and the conditional mean of the cluster-level outcome: $Q= (Q_{E,{\bf W}},\bar{Q}^c)$. Now, however,  the conditional mean of the cluster-level outcome is assumed to be an average of common individual-level regressions: $\bar{Q}_0^c(A,E,\mathbf{W})= \sum_i   \alpha_{i\LargerCdot} \bar{Q}_0(A,E,W_{i\LargerCdot} )$.

The  efficient influence curve of $\Psi^{II}$ at $\mathbb{P}_0\in \mathcal{M}^{II}$ is given by
\begin{equation}
D^{II}(\mathbb{P}_0)(O)=
\sum_{i=1}^N  \alpha_{i\LargerCdot}
\left(  \frac{\mathbb{I}(A=a)}{g_0(A\mid E,W_{i\LargerCdot})}\big(Y_{i\LargerCdot}-\bar{Q}_0(A,E,W_{i\LargerCdot})\big) +
\bar{Q}_0(a,E,W_{i\LargerCdot})-\Psi^{II}(\mathbb{P}_0)(a) \right).
\label{effic2}
\end{equation}
Under  sub-model $\mathcal{M}^{II}$, the efficient influence curve 
is the average of an individual-level function. The  first component of this individual-level function is  the weighted deviations between the individual-level outcome and its expectation given the exposure and covariates; the weight is the inverse of the individual-level propensity score. The second component is the deviation between the conditional expectation of the individual-level outcome and the target parameter. 

As before, the efficient influence curve (Eq.~\ref{effic2}) is used to derive the TMLE. Specifically, suppose we  have an initial estimator $\hat{\bar{Q}}(A, E, {W})$ of the individual-level outcome regression $\bar{Q}_0(A, E, {W})$ and an estimator $\hat{g}(a | E, {W})$ of the individual-level propensity score $g_0(a | E, {W})$.
The TMLE algorithm updates the initial estimator $\hat{\bar{Q}}(A, E, {W})$ into $\hat{\bar{Q}}^*(A, E, {W})$ by minimizing a pre-specified loss function along a least favorable (with respect to the statistical estimand) sub-model through $\hat{\bar{Q}}(A, E, {W})$. This updating step also serves to target the initial cluster-level outcome regression $\hat{\bar{Q}}^{c}= \sum_i  \alpha_{i\LargerCdot}  \hat{\bar{Q}}$ into $\hat{\bar{Q}}^{c*} =  \sum_i  \alpha_{i\LargerCdot}  \hat{\bar{Q}}^*$.

As loss function for the outcome regression, we use  the average of an $i$-specific loss function:
\begin{equation}\label{loss_gen}
\mathcal{L}^{II}(\bar{Q}^c)(O)= \sum_{i=1}^N  \alpha_{i\LargerCdot} \mathcal{L}(\bar{Q})(O)
,\end{equation}
where 
\begin{equation}
- \mathcal{L}(\bar{Q})(O)=Y_{i\LargerCdot}\log \big[\bar{Q}(A,E,W_{i\LargerCdot})\big]+(1-Y_{i\LargerCdot})\log \big[1-\bar{Q}(A,E,W_{i\LargerCdot})\big]. 
\end{equation}
$\mathcal{L}$ is  a valid loss function for the $i$-specific outcome regression $\mathbb{E}_0(Y_{i\LargerCdot} | A,E,W_{i\LargerCdot})$, and under the sub-model $\mathcal{M}^{II}$ this regression is constant across individuals $\bar{Q}_0(A,E,W)$. Therefore, this is a valid loss function for each $i$, and the sum loss is also valid (Appendix C).
%

For our fluctuation model through an initial estimator $\hat{\bar{Q}}(A,E,W)$, we select the individual-level analog to the cluster-level fluctuation model (Eq.~\ref{fluc_c}):
\begin{equation} 
logit\big[ \hat{\bar{Q}}(\epsilon) \big]= logit\big[\hat{\bar{Q}}\big]+\epsilon \hat{H},\end{equation}
where the individual-level clever covariate is defined as 
\begin{equation}
\hat{H}_{ij} =\frac{\mathbb{I}(A_j=a)}{\hat{g}( a\mid E_j, W_{ij})},  \ i=1,\ldots,N_j \text{ and } j=1,\ldots, J.
\end{equation}
This fluctuation model is only a function of the covariate matrix $\mathbf{W}$ through the $i^{th}$-specific covariate $W_{i\LargerCdot}$ and is  a sub-model of $\mathcal{M}^{II}$.
At zero fluctuation, the initial estimator is returned. 
This combination of loss function and fluctuation model has score $\frac{d}{d\epsilon}\mathcal{L}^{II}(\bar{Q}^c)(\epsilon)$ at $\epsilon=0$  that spans  the relevant portion of the efficient influence curve $D^{II}$.

The amount of fluctuation $\epsilon$ is fit 
by  pooling individuals across clusters and running  logistic regression of the individual-level outcome $Y_{i\LargerCdot}$ on the clever covariate $\hat{H}_{i\LargerCdot}$ with the $logit$ of the initial estimator $\hat{\bar{Q}}(A, E, W_{i\LargerCdot})$ as offset and  weights $\alpha_{i\LargerCdot}$. 
Plugging in the resulting coefficient estimate $\hat{\epsilon}$ provides an updated fit of the individual-level regression 
\begin{equation}
\hat{\bar{Q}}^*=\hat{\bar{Q}}(\hat{\epsilon}) = logit^{-1}\big[ logit(\hat{\bar{Q}})+\hat{\epsilon}  \hat{H}\big]
\end{equation}
and thereby the cluster-level regression: $\hat{\bar{Q}}^{c*}(A,E, \mathbf{W}) =  \sum_{i=1}^N  \alpha_{i\LargerCdot}  \hat{\bar{Q}}^*(A,E,W_{i\LargerCdot})$.

As an initial estimator of the covariate distribution, we again use the empirical distribution $\hat{Q}_{E,\mathbf{W}}$, which puts weight $1/J$ on each cluster.  
 As before, the empirical distribution $\hat{Q}_{E,\mathbf{W}}$ is the non-parametric maximum likelihood estimator and 
  does not need to be targeted.\citep{Chpt5, vanderLaan2014bigdata}   Therefore, the TMLE 
is  defined as the substitution estimator obtained by plugging $\hat{Q}^*=(\hat{Q}_{E,\mathbf{W}},\hat{\bar{Q}}^{c*})$ into the parameter mapping $\Psi^{II}$:
\begin{equation}
\Psi^{II}(\hat{Q}^*)(a)=  \frac{1}{J}\sum_{j=1}^J \left\{ \sum_{i=1}^{N_j}  \alpha_{ij}  \hat{\bar{Q}}^*(a,E_j,W_{ij}) \right\} =\frac{1}{J}\sum_{j=1}^J \hat{\bar{Q}}^{c*}(a,E_j,{\bf W}_j) .
\end{equation}
The point estimate, denoted $\hat{\psi}^{II}(a)$,  is the sample average of the targeted predictions of the cluster-level outcome, given the exposure of interest $(A=a)$ and the measured covariates.
By construction, this TMLE solves the efficient influence curve equation:
 $\sum_j D^{II}(\hat{Q}^*,\hat{g})(O_j)=0$. Thereby, the estimator is double robust and asymptotically efficient under consistent estimation of both the outcome regression and propensity score.
 
 Statistical inference proceeds as presented in Section \ref{ClusterLevelTMLE}. 
Specifically, let 
 \begin{equation}
 \hat{D}^{II}(\hat{Q}^*, \hat{g})(O_j) = \sum_{i=1}^{N_j}  \alpha_{ij} 
 \left( \frac{\mathbb{I}(A_j=a)}{\hat{g}(A_j\mid E_j,W_{ij})}\big(Y_{ij}-\hat{\bar{Q}}^{*}(A_j,E_j,W_{ij})\big) +
\hat{\bar{Q}}^*(a,E_j,W_{ij})-\hat{\psi}^{II}(a) \right)
\end{equation}
  be the plug-in estimator of the  influence curve  for observation $O_j$. 
 We obtain a variance estimator with the sample variance of $\hat{D}^{II}(\hat{Q}^*, \hat{g})$ divided by the number of experimental units: $\hat{\sigma}^2=Var[\hat{D}^{II}]/J$. 
 For this  sub-model, an alternative variance estimator, which explicitly estimates the correlation structure within each cluster, is proposed in Schnitzer \emph{et al.}. \citep{Schnitzer2014}
 
 Appendix D provides step-by-step implementation of the individual-level TMLE and corresponding \texttt{R} code.  This individual-level TMLE can also be implemented with the existing  \texttt{ltmle}\citep{ltmlePackage} package using  \texttt{id} to  specify the clusters (independent units) and \texttt{observation.weights} for the weights $\alpha_{ij}$. It is worth emphasizing, however, that this individual-level TMLE for the impact of a cluster-level exposure is developed under a causal model with strong assumptions (Eq.~\ref{SCM2}).  In the following Sections we explore the theoretical and practical consequences of these assumptions.

\section{Theoretical comparison of the TMLEs}

The cluster-level TMLE is derived under a general causal model allowing for arbitrary dependence of individuals within a cluster. Our contribution is to propose incorporating pooled individual-level regressions as candidates in the Super Learner  library for initial   estimation of the expected cluster-level outcome $\bar{Q}_{0}^c(a,E, \mathbf{W})$.
In contrast,  the individual-level TMLE is derived under the restricted causal model,  which assumes that the covariates of one individual do not affect the outcome of another (i.e. no covariate interference) and that  the cluster-level covariates $E$ and individual $i$-specific covariates $W_{i\LargerCdot}$ are sufficient to control for confounding.
In practice, implementation of the two estimators differs in where and when we take averages. In the larger model, we immediately average any individual-level regressions to obtain an initial estimator $\hat{\bar{Q}}^c(a, E, \mathbf{W})$ and target using a cluster-level  clever covariate. In the sub-model, we update the individual-level estimator  $\hat{\bar{Q}}(a,E, {W})$ using an individual-level clever covariate  and then average the targeted predictions within each cluster: 
$\hat{\bar{Q}}^{c*}(a,E, \mathbf{W}) =  \sum_i \alpha_{i\LargerCdot} \hat{\bar{Q}}^*(a,E, W_{i\LargerCdot})$.

 To compare the asymptotic efficiency of the two approaches, we first consider the special case where the exposure assignment $A$ is independent of the whole covariate matrix $\mathbf{W}$, given the 
 environmental factors $E$. In the running example, this would hold by design if the Test-and-Treat intervention were randomized  $g_0^c(A | E, \mathbf{W}) = 0.5$. More generally, this condition would hold if the  intervention were rolled out according only to community-level characteristics, such as baseline HIV prevalence and perceived need: $g_0^c(A | E, \mathbf{W}) = g_0^c(A | E)$. In this case, the efficiency bound for $\Psi^I(\mathbb{P}_0)$, presented in Section~\ref{ClusterLevelTMLE}, will be identical to the efficiency bound for $\Psi^{II}(\mathbb{P}_0)$, presented in Section~\ref{Sect:CausalSubmodel}. In other words, we have the efficient influence curves are equal: $D^{I}(\mathbb{P}_0)=D^{II}(\mathbb{P}_0)$ at a $\mathbb{P}_0\in \mathcal{M}^{II}$ (Proof in Appendix E).
 
However,  this does not imply that the corresponding TMLEs will be identical if the propensity score is unknown. In an observational setting, estimating a cluster-level propensity score $g_0^c(a | E, \mathbf{W})$ when implementing the TMLE for $\Psi^I(\mathbb{P}_0)$ as compared to an individual-level propensity score $g_0(a | E, {W})$ when implementing the TMLE for $\Psi^{II}(\mathbb{P}_0)$ can result in estimators that are asymptotically distinct. 
 If on the other hand,  the exposure mechanism depends on both the environmental factors  and the covariate matrix $g_0^c(A | E, \mathbf{W})\ne g_0^c(A | E)$, then the efficiency bound for $\Psi^{II}(\mathbb{P}_0)$ in the smaller model $\mathcal{M}^{II}$ will be better than the efficiency bound for $\Psi^I(\mathbb{P}_0)$ in the larger model $\mathcal{M}^I$. 

\section{Finite sample simulations}

In this section, we investigate the practical performance of the two TMLEs. We begin with a simple simulation to demonstrate implementation and performance in an observational setting. We then present a more realistic simulation, generated to reflect the  HIV prevention and treatment example. Throughout, the causal parameter is the population average treatment effect
$\mathbb{E}[Y^c(1) - Y^c(0)]$.
 All simulations were conducted using \texttt{R}.\citep{R} Full computing code is publicly available. 

\subsection{Simulation 1 - Simple observational setting}

We consider a sample size of $J=100$ clusters. For each unit $j=\{1, \ldots, J\}$, we draw the number of individuals $N_j$ from a normal with mean 50 and standard deviation 10 and round to the nearest whole number. 
Then for each individual $i=\{1,\ldots, N_j\}$, two covariates $(W1, W2)$ are drawn from a multivariate normal. 
We include their averages as cluster-level covariates:  $W1_j^c = 1/N_j \sum_{i} W1_{ij}$ and  $W2_j^c = 1/N_j \sum_{i} W2_{ij}$. 
We consider
an observational setting where the propensity score depends on one cluster-level aggregate:
$A_j \sim Bern( logit^{-1}(0.75 W1^c_j))$.
The probability of the individual-level outcome is simulated under two data generating distributions. Specifically, we vary the strength of the coefficients to simulate scenarios
with minimal covariate interference 
\begin{equation}\label{SimWorkModelHolds}
\mathbb{P}_0(Y_{ij}=1\mid A_j, W1^c_j, W2^c_j, W1_{ij}, W2_{ij}) = logit^{-1}(0.25 + 0.1A_j + 0.15W1^c_j + 1.15W1_{ij} + W2_{ij})
 \end{equation}
and with stronger covariate interference
\begin{equation}\label{SimWorkModelFails}
\mathbb{P}_0(Y_{ij}=1 \mid A_j, W1^c_j, W2^c_j, W1_{ij}, W2_{ij}) =  logit^{-1}(0.25 + 0.1A_j +  0.15W1^c_j +  0.25W1_{ij} + W2^c_j).
 \end{equation}
In the first data generating process (Eq.~\ref{SimWorkModelHolds}), an individual's outcome is strongly impacted by his/her own covariates ($W1_{ij}$, $W2_{ij}$) and only weakly impacted by the covariates of others (i.e. $W1^c_j$). In the second process (Eq.~\ref{SimWorkModelFails}), the opposite  holds.
We then simulate the binary individual-level outcome as
\begin{equation}
Y_{ij} = \mathbb{I}\big(U_{Y_{ij}} < \mathbb{P}_0(Y_{ij} \mid A_j, W1^c_j, W2^c_j, W1_{ij}, W2_{ij}) \big).
 \end{equation}
where the unmeasured error $U_{Y_{ij}} \in [0,1]$ is generated under two scenarios: independent within a cluster and correlated  within a cluster. In the former,  $U_{\mathbf{Y}_{j}}=(U_{Y_{ij}}: i, \ldots, N_j)$ is generated by  independently drawing $N_j$ times from a Uniform(0,1), while in the latter $U_{\mathbf{Y}_{j}}$ is generated by applying the cumulative distribution function to correlated normal random variables (Full \texttt{R} code in Appendix D).  Varying the dependence of the unmeasured factors  $U_{\mathbf{Y}}$ determining the outcomes $\mathbf{Y}$ within a cluster allows us to examine the randomization assumption inherent in the restricted causal model (Eq.~\ref{SCM2}). 
In practice, independent $U_\mathbf{Y}$ might be reasonable for outcomes that are not biologically or socially transmitted, but may be unreasonable otherwise (Supplementary Figure~S2).

As before, we define the cluster-specific outcome $Y^c$ as the empirical mean of the individual-level outcomes within that cluster. 
We  generate  counterfactual outcomes $(Y^c(1), Y^c(0))$ by setting the cluster-level exposure to $A=1$ and $A=0$, respectively. For each data generating process, the average treatment effect is calculated by taking the mean difference in the counterfactual cluster-level outcomes for a population of 10,000 clusters (Supplementary Table~S1). 
We also simulate under the null by setting the counterfactual outcome under the intervention equal to counterfactual outcome under the control. 

As shown in Table~\ref{Table:Sim1Estimators}, we consider three targeted estimators: TMLE-$Ia$ adjusting for the covariates at the cluster-level in both the outcome regression and the propensity score regression; TMLE-$Ib$ adjusting at the individual-level  in the outcome regression and at the cluster-level in the propensity score regression; and TMLE-$II$ adjusting at the individual-level in both the outcome regression and propensity score regression. TMLE-$Ia$ and TMLE-$Ib$ correspond to statistical model $\mathcal{M}^{I}$ and TMLE-$II$ to  sub-model $\mathcal{M}^{II}$. Both TMLE-$Ib$ and TMLE-$II$ harness the pairing of individual-level covariates and outcomes, but the former incorporates this information as working assumptions (Eq.~\ref{workingmodelindiv})  during the estimation step , while the latter assumes  the restricted causal model  (Eq.~\ref{SCM2}) reflects the true data generating process.
We compare the targeted estimators to the  unadjusted estimator, the average difference in cluster-level outcomes between treated and control groups.

\begin{table}[!htbp]
\small\sf\centering
\caption{Targeted estimators considered for Simulation 1: ``Causal Model" refers to the causal model assumed during development of the estimator. ``Cluster-level" refers to logistic regression after all the data are aggregated. ``Individual-level" refers to logistic regression pooling individuals across clusters and with weights $\alpha_{ij}=1/N_j$. 
}
\begin{tabular}{  l l l l l   }
\toprule
Estimator & Causal Model   & Outcome Regression  &  Pscore Regression & Targeting  \\
 \midrule
  TMLE-$Ia$  &  General (Eq.~2.1) & Cluster-level & Cluster-level   & Cluster-level  \\\
  TMLE-$Ib^*$  &  General (Eq.~2.1)		 & Individual-level$^{**}$ $\quad$ & Cluster-level & Cluster-level  \  \\
TMLE-$II$ & Restrictive (Eq.~4.1) & Individual-level & Individual-level & Individual-level   \\
\bottomrule
\multicolumn{5}{l}{$^*$During estimation consider working assumptions  to generate alternative estimators of $\bar{Q}_0^c$.} \\
\multicolumn{5}{l}{$^{**}$Run a pooled individual-level regression \& then average individual-level predictions within clusters.}
\end{tabular}
\label{Table:Sim1Estimators}
\end{table}

\begin{table}
\small\sf\centering
\caption{Estimator performance in Simulation 1 under minimal covariate interference (Eq.~\ref{SimWorkModelHolds})  and under stronger covariate interference  (Eq.~\ref{SimWorkModelFails}). We also vary the dependence of the unmeasured factors determining the outcome $U_\mathbf{Y}$: independent (top) and correlated (bottom). 
Performance is given by 
bias as the average deviation between the estimate and truth; $\sigma$ as the standard error; rMSE as the root-mean squared error; power as the proportion of times the false null hypothesis is rejected, and coverage as the proportion of times the 95\% confidence interval contains the true value.
 All measures are in \%. }
\begin{tabular}{l lllll lllll }
  \toprule
              & \multicolumn{5}{l}{Minimal covariate interference}  &  \multicolumn{5}{l}{Stronger covariate interference }  \\
Estimator &   Bias & $\sigma$& rMSE  & Power & Coverage  &    Bias & $\sigma$ & rMSE & Power & Coverage  \\
\midrule
Unadj. & 10.4 & 5.0 & 11.5 & 66 & 46 & 7.6 & 3.8 & 8.5 & 72 & 49 \\ 
  TMLE-$Ia$ & 0.0 & 1.2 & 1.2 & 28 & 95 & 0.0 & 1.4 & 1.4 & 34 & 94 \\ 
  TMLE-$Ib$ & 0.0 & 1.2 & 1.2 & 27 & 95 & 0.0 & 1.4 & 1.4 & 23 & 98 \\ 
  TMLE-$II$ & 0.2 & 1.2 & 1.2 & 34 & 95 & 1.7 & 1.6 & 2.3 & 65 & 81 \\ 
    & \multicolumn{8}{c}{Independent $U_{\mathbf{Y}}$ determining the outcome } \\
\midrule
Unadj. & 6.3 & 3.2 & 7.1 & 88 & 48 & -3.6 & 2.4 & 4.3 & 21 & 67 \\ 
  TMLE-$Ia$ & -0.0 & 1.3 & 1.3 & 86 & 94 & 0.0 & 1.7 & 1.7 & 96 & 94 \\ 
  TMLE-$Ib$ & -0.0 & 1.3 & 1.3 & 28 & 100 & 0.0 & 1.8 & 1.8 & 91 & 98 \\ 
  TMLE-$II$ & -4.1 & 2.4 & 4.7 & 5 & 58 & -2.1 & 2.0 & 3.0 & 56 & 81 \\ 
   & \multicolumn{8}{c}{Dependent $U_{\mathbf{Y}}$ determining the outcome } \\
\bottomrule
\end{tabular}
\label{Table:Sim1Results}
\end{table}

\subsubsection{Results:}

Table~\ref{Table:Sim1Results} provides a summary of the estimator performance over 5,000 repetitions of the simulation. Recall the unadjusted estimator is simple the difference in average outcomes among treated units and average outcomes among control units. TMLE-$Ia$, developed under the general model $\mathcal{M}^I$, corresponds to an aggregated data approach;  cluster-level regressions are used for  both initial estimation and targeting. TMLE-$Ib$, also developed under the general model $\mathcal{M}^I$, uses a pooled  individual-level regression for initial estimation of the mean outcome and then a cluster-level regression for updating.  TMLE-$II$, developed under the more restrictive sub-model $\mathcal{M}^{II}$, uses pooled individual-level regressions for both initial estimation and updating.

When the unmeasured factors determining the outcome $U_\mathbf{Y}$ are independent,
the unadjusted estimator, which fails to control for measured confounding, is biased. This bias is substantial enough to prevent reliable inference; the 95\% confidence interval coverage is $<$50\%.
The TMLE corresponding to an aggregated data approach  (TMLE-$Ia$) performs well with negligible bias and good confidence interval coverage. 
 When there is minimal covariate interference  (Eq.~\ref{SimWorkModelHolds}), TMLE-$Ib$, which makes working assumptions for initial estimation of the outcome regression, 
  performs similarly to  TMLE-$Ia$.
 However, when there is stronger covariate interference (Eq.~\ref{SimWorkModelFails}) and these working assumptions  fail,  TMLE-$Ib$ provides less power (23\% vs. 34\%) and conservative confidence interval coverage (98\%). 
 In this scenario,  the cluster-level regression provides a better  approximation of the true outcome regression resulting in greater efficiency and power for the aggregated estimator (TMLE-$Ia$).
 
 Under independent errors and minimal covariate interference (Eq.~\ref{SimWorkModelHolds}), TMLE-$II$, constructed under the restricted causal model, performs well with good confidence interval coverage and results in notably more power (34\%).  However, with stronger covariate interference (Eq.~\ref{SimWorkModelFails}),  TMLE-$II$ is  biased and provides misleading inference. 
 Its confidence interval coverage is much less than nominal (81\%), while deceivingly providing the most power (65\%).  Under the null, we also see inflated Type I error rates of 18\% (Supplementary Table~S2). 

When the unmeasured factors determining the outcome $U_\mathbf{Y}$ are correlated, the assumptions in the restricted causal model (Eq.~\ref{SCM2})  do not hold (Supplementary Figure~S2b). As expected, the unadjusted estimator is again biased with 95\% confidence interval coverage ranging from 48\% to 67\%. Both targeted estimators developed under the general model (TMLE-$Ia$ and TMLE-$Ib$) have negligible bias, but the cluster-level estimator yields more power. The individual-level TMLE developed under the restricted model (TMLE-$II$) now exhibits substantial bias regardless of the strength of covariate interference. Its resulting confidence interval coverage is much less than the nominal  and type I error reaches $>$40\% (Supplementary Table~S2). 


In summary, 
when the assumptions in the more restrictive causal model  hold, the individual-level targeted estimator (TMLE-$II$) is the most powerful. 
However, if these commonly made assumptions fail, this TMLE is biased and can yield misleading inference in an observational setting.
Incorporating  working assumptions during the estimation stage (Eq.~\ref{workingmodelindiv}) is more robust than assuming they hold in the underlying causal model (Eq.~\ref{SCM2}). 
Specifically, TMLE-$Ib$ provides a mechanism to leverage the pairing of individual-level covariates and outcomes, while avoiding additional causal assumptions. 
 In practice, we recommend  considering a general TMLE-$I$ which includes both cluster-level  and individual-level specifications in the Super Learner library for initial estimation of the outcome regression. This TMLE is implemented in the following simulation study.


\subsection{Simulation 2 - HIV prevention and treatment trial}

We now consider a more complicated simulation, generated to reflect the running example. 
For 1000 iterations, we simulate a cluster randomized Test-and-Treat trial, 
 consisting of 32 communities with 200 individuals each. 
 Within each community,  we generate an underlying sexual network through a degree-corrected, bipartite stochastic block model.\citep{Karrer2011}
On each network, we  simulate an HIV epidemic with a susceptible-infected-recovered compartmental model.\citep{Anderson1992} 
In the intervention arm, 85\% of the HIV-positive patients are on ART and have successfully suppressed viral replication. In the control arm, 55\% of the HIV-positive patients are on ART and are suppressed.\citep{WHO2015guidelines,BCPP909090, Fidler2016CROI,Petersen2016CascadeIAS}
There is no sexual mixing or spillover effects across communities. 
To initiate the epidemic in each community, we randomly select 10\% of individuals to be infected and allow the virus to spread until an average prevalence of 25\% is reached. We then begin the study and follow all communities for three years. %
Full Python code to generate the networks and epidemic is available in Staples.\citep{Staples2016Code} 

As before, the target of inference is the population average treatment effect: the expected difference in the counterfactual cumulative HIV incidence under the Test-and-Treat intervention and under the standard of care. Within each community, 75 baseline HIV-negative individuals are selected, and   
the cluster-level outcome is the proportion who seroconvert within the three years of follow-up. 
The true value of the  treatment effect is calculated by averaging the difference in the cluster-level counterfactual outcomes in the  population of all clusters from all trials ($32 \times 1000$). 
The estimated impact of the Test-and-Treat intervention is -4.0\%, reducing HIV incidence from 9.1\% under the standard of care to 5.1\% under the intervention. 
We also simulate under the null by setting the counterfactual outcome under the intervention equal to the counterfactual outcome under the control.

%

We  consider the following individual-level adjustment variables: demographic risk group,  degree (number of sexual partners), and number of partners infected at baseline. We also consider the following cluster-level adjustment variables: 
baseline HIV prevalence, assortativity (degree-degree correlation across all network connections), and number of components (number of distinct sexual groups). 
To select among candidate adjustment variables,
we apply  a discrete Super Learner to data-adaptively select the candidate TMLE, which minimizes variance and maximizes precision.\citep{Balzer2016DataAdapt} This procedure incorporates ``collaborative" \citep{Gruber2010ctmle} estimation of the known propensity score $g_0^c(A | E, \mathbf{W}) = g_0(A | E, {W}) = 0.5$ for further gains in precision. 

We implement this approach  under the larger general model (TMLE-$I$) and under the smaller sub-model (TMLE-${II}$). Both TMLEs include pooled individual-level regressions as candidate estimators of the conditional mean of the cluster-level outcome $\bar{Q}_0^c(A, E, \mathbf{W})$. The former estimates the propensity score and targets at the cluster-level, while the latter estimates the propensity score and targets at the individual-level. In other words, TMLE-$I$ can be considered a hybrid of TMLE-$Ia$ (aggregated data approach) and TMLE-$Ib$ (incorporating working assumptions), which were studied in the previous section. In this simulation, the restricted causal model (Eq.~\ref{SCM2}) does not hold due to causal interactions between individuals within a community (i.e. sexual transmission of HIV through the network). Nonetheless, the finite sample performances of the TMLEs is expected to be similar due to randomization of the exposure (i.e. the double robustness property). We compare the targeted approaches to the unadjusted estimator,  inverse probability of treatment weighting  (IPTW) adjusting for average degree in the propensity score regression, and G-computation adjusting for average degree in the outcome regression. 

\subsubsection{Results:}

\begin{table}[!h]
\small\sf\centering
\caption{Estimator performance in Simulation 2 when there is an effect and under the null. 
Performance is measured by bias as the average deviation between the estimate and truth;
 ${\sigma}$ as the standard error; 
rMSE as the root-mean squared error; power as the proportion of times the false null hypothesis is rejected; coverage as the proportion of times the 95\% confidence interval contains the true value, and Type I error as the proportion of times the true null hypothesis is rejected.
 All measures are in \%.}
\begin{tabular}{l   ccccc  ccccc}
\toprule
	& \multicolumn{5}{c}{\textbf{With an effect}} &  \multicolumn{5}{c}{\textbf{Under the null}} \\
 &   Bias & $\sigma$ & rMSE & Power & Coverage & Bias & $\sigma$ & rMSE & Type I & Coverage \\ 
  \midrule
Unadj. & 0.1 & 1.6 & 1.6 & 66 & 94 & 0.1 & 1.9 & 1.9 & 4 & 96 \\ 
  IPTW & 0.1 & 1.6 & 1.6 & 68 & 96 & 0.0 & 1.8 & 1.8 & 3 & 97 \\ 
  Gcomp. & 0.1 & 1.5 & 1.6 & 75 & 93 & 0.0 & 1.8 & 1.8 & 6 & 94 \\ 
  TMLE-$I$ & 0.1 & 1.3 & 1.3 & 83 & 96 & 0.0 & 1.5 & 1.5 & 5 & 95 \\ 
  TMLE-$II$ & 0.1 & 1.3 & 1.3 & 82 & 95 & 0.0 & 1.5 & 1.5 & 5 & 95 \\ 
\bottomrule
\end{tabular}
\label{Table:Sim2Results}
\end{table}

As expected, all estimators are unbiased and adjustment for baseline covariates increases precision and power in this trial setting (Table~\ref{Table:Sim2Results}).\citep{Fisher1932, Cochran1957, Cox1982, Tsiatis2008, Moore2009, Rosenblum2010, Balzer2016DataAdapt} The unadjusted difference in cluster-level mean outcomes yields 66\% power, while IPTW yields 68\%, and parametric G-computation yields 75\%. The two TMLEs, data-adaptively adjusting for the covariate(s) to increase precision, obtain substantially more power (82-83\%), while maintaining nominal confidence interval coverage and Type I error control. In both TMLEs, the number of partners infected at baseline (an individual-level covariate) is selected as the adjustment variable for the outcome regression in 76\% of the trials (Supplementary Table S3). 
The slight difference in performance between the two TMLEs is due to targeting, which occurs at the cluster-level in TMLE-$I$ and at the individual-level in TMLE-$II$. 
Overall, these simulations demonstrate that in a trial setting, 
the utility of the working assumption (Eq.~\ref{workingmodelindiv})  is wider flexibility in covariate adjustment to increase efficiency without creating bias.

\section{Application - Household socioeconomic status and baseline HIV testing in SEARCH}

The  Sustainable East Africa Research in Community Health Study (SEARCH) is an ongoing cluster randomized trial to evaluate the impact of a community-based strategy for early HIV diagnosis with immediate and streamlined ART on HIV incidence in rural Uganda and Kenya (NCT:01864603).
In SEARCH, population-based HIV testing was conducted through  multi-disease community health campaigns, consisting of out-of-facility health fairs followed by home-based testing for non-attendees. \cite{Chamie2016} 
HIV testing was successfully completed for 89\% (131,307/146,906) of residents who were aged $\geq15$ years and considered stable ($\geq$ 6 months in the community during the past year) at baseline.  
 Since data collection for the primary outcome  is ongoing, we apply the proposed methods to estimate the association of household socioeconomic status on the risk of not testing for HIV at baseline. 

In this application, the  cluster is the household, and the cluster-based exposure is an indicator of living in a household in the lowest socioeconomic class, calculated using principal component analysis of ownership of livestock and household items.\citep{Chamie2016} The individual-level outcome is an indicator of failing to test for HIV, and the cluster-level outcome is the proportion of adults not testing in a given household. 
The cluster-level confounders include community indicators, the size of the household, and an indicator of male head of household (Table~\ref{Table:App}).  The individual-level confounders include age, sex, educational attainment, occupation type, marital status, and mobility (indicator of living 1 or more months away from the community).   The target parameter is the standardized  risk difference, corresponding to the causal risk difference if the necessary assumptions hold. 

\begin{table}[!htbp]
\small\sf\centering
\caption{Characteristics of baseline adult 
 residents of the 16 SEARCH intervention communities (5 in Eastern Uganda, 5 in Southwestern Uganda, and 6 in Kenya) with complete socioeconomic  information (249 individuals excluded). Analyses also adjusted for community indicators.}
\begin{tabular}{l llll}
  \toprule
 & E. Uganda & S.W. Uganda & Kenya & Overall \\ 
  \midrule
N. individuals & 25041 & 24913 & 27571 & 77525 \\ 
N. households & 10106 & 9939 & 11979 & 32024 \\ 
 Male & 11365 (45\%) & 11641 (47\%) & 12137 (44\%) & 35143 (45\%) \\ 
Age in years & \\
\quad  15-24 & 9572 (38\%) & 8466 (34\%) & 9226 (33\%) & 27264 (35\%) \\ 
\quad  25-34 & 5305 (21\%) & 5709 (23\%) & 6669 (24\%) & 17683 (23\%) \\ 
 \quad 35-44 & 3986 (16\%) & 4363 (18\%) & 4235 (15\%) & 12584 (16\%) \\ 
 \quad 45+ & 6178 (25\%) & 6375 (26\%) & 7441 (27\%) & 19994 (26\%) \\ 
 Education & \\
\quad  Less than primary & 3855 (15\%) & 4413 (18\%) & 2132 (8\%) & 10400 (13\%) \\ 
\quad  Primary & 15255 (61\%) & 13966 (56\%) & 22302 (81\%) & 51523 (66\%) \\ 
\quad  Secondary or higher & 5931 (24\%) & 6534 (26\%) & 3137 (11\%) & 15602 (20\%) \\ 
  Occupation & \\
\quad  Formal$^a$  & 5826 (23\%) & 5273 (21\%) & 6604 (24\%) & 17703 (23\%) \\ 
\quad  High risk informal$^b$ & 397 (2\%) & 652 (3\%) & 2331 (8\%) & 3380 (4\%) \\ 
\quad  Low risk informal$^c$ & 17190 (69\%) & 16318 (65\%) & 15361 (56\%) & 48869 (63\%) \\ 
\quad  Jobless  or disabled & 751 (3\%) & 1132 (5\%) & 2066 (7\%) & 3949 (5\%) \\ 
\quad    Other & 877 (4\%) & 1538 (6\%) & 1209 (4\%) & 3624 (5\%) \\ 
  Never married & 6913 (28\%) & 7424 (30\%) & 7515 (27\%) & 21852 (28\%) \\ 
Mobile$^d$ & 3024 (12\%) & 3305 (13\%) & 1960 (7\%) & 8289 (11\%) \\ 
 Male household head & 18219 (73\%) & 16247 (65\%) & 16120 (58\%) & 50586 (65\%) \\ 
Household size$^e$ & 3 (2, 4) & 3 (2, 4) & 3 (2, 4) & 3 (2, 4) \\ 
Lowest SES$^f$ & 4201 (17\%) & 5212 (21\%) & 2522 (9\%) & 11935 (15\%) \\ 
Did not test for HIV& 2434 (10\%) & 2604 (10\%) & 3439 (12\%) & 8477 (11\%) \\ 
   \bottomrule
 \multicolumn{5}{l}{\footnotesize $^a$Formal: teacher, student, government worker, military worker, health worker, factory worker} \\
  \multicolumn{5}{l}{\footnotesize $^b$High risk informal: fishmonger, fisherman, bar owner, bar worker, transport, tourism} \\
 \multicolumn{5}{l}{\footnotesize $^c$Low risk informal: farmer, shopkeeper, market vendor, hotel worker, housewife, household}\\
 \multicolumn{5}{l}{\footnotesize worker, construction worker, mining} \\
  \multicolumn{5}{l}{\footnotesize $^d$Mobile: $\geq$ 1 month/past year away from the community} \\
      \multicolumn{5}{l}{\footnotesize $^e$Median with interquartile range} \\
    \multicolumn{5}{l}{\footnotesize $^f$Lowest SES: Living in a household with the lowest quintile of the  wealth index} \\
\end{tabular}
\label{Table:App}
\end{table}

In this setting, we are willing to assume that after controlling for the cluster-level confounders and exposure, each individual's outcome is not a direct function of other household members' individual-level covariates. We are also willing to assume that the conditional expectation of the individual-level outcome is common across individuals. Therefore, under the general model $\mathcal{M}^I$ and using the working assumptions in Eq.~\ref{workingmodelindiv}, we implement TMLE with Super Learner to fully leverage the pairing of individual-level risk factors and outcomes, while avoiding unwarranted assumptions (\texttt{SuperLearner}-v2.0-21\citep{SuperLearnerPackage}). 
 The  library of candidate algorithms includes both parametric and semi-parametric approaches: main terms logistic regression without and without all possible pairwise interactions, generalized additive models (\texttt{gam}-v1.14\citep{gamR}), and penalized maximum likelihood (\texttt{glmnet}-v2.0-5\citep{glmnetR}). We use the same library for estimation of the outcome regression and the propensity score.  The analysis is restricted to the 16 intervention communities (77,525 adults total), and the household is  the unit of independence: $J$=32,024.

After controlling for measured confounders, the marginal risk of not testing associated with living in household in the lowest socioeconomic class  is 10.7\%, while the marginal risk of not testing associated with living in a household in a higher  socioeconomic class is 10.0\%. Despite the large sample size, the standardized risk difference of 0.7\% (95\%CI: -0.1\%, 1.4\%) is not significant at the 0.05-level. For comparison, the unadjusted estimator, which fails to control for  confounding,  yielded a risk difference of  -0.3\% (-1.0\%, 0.3\%). 

\section{Concluding remarks}\label{sectconclude}

In this manuscript, we  present two distinct approaches for leveraging a hierarchical data structure to improve the performance of double robust TMLEs for the causal effect of a cluster-level exposure. The first assumes a general hierarchical causal model, which allows for arbitrary dependence  of individuals within clusters. 
For the corresponding  statistical model $\mathcal{M}^I$, we review a cluster-level TMLE, which is a direct analog for the individual-level TMLE in non-hierarchical setting. Our novel contribution to this cluster-level estimator is to use the pairing of individual-level covariates and outcomes for improved estimation of the expected cluster-level outcome. 
Pooled individual-level regressions can lead to both asymptotic and finite sample improvements without placing restrictions on the original statistical model. 
%
Super Learner provides one way to choose between and combine several candidate algorithms, including cluster-level parametric regressions, averages of individual-level regressions, and more data-adaptive methods. 

We then consider a more restrictive  causal sub-model, which assumes that the cluster-level and individual $i$-specific covariates are sufficient to control for confounding. 
 For the corresponding restricted statistical model $\mathcal{M}^{II}$, we present an alternative individual-level TMLE, which still targets the relevant cluster-level causal effect. 
When the assumptions in the sub-model hold, this TMLE  is guaranteed asymptotically to be at least as efficient as the TMLE developed under the general causal model.
When the assumptions  fail, this TMLE  may be subject to bias and misleading inference in an observational setting. However, if the propensity score is consistently estimated, the individual-level TMLE 
will remain consistent due to its double robustness property,  representing an important advantage over alternative estimators, such as those based on a single regression (e.g. IPTW and G-computation). 

%

The results of this paper  have the following practical implications. 
When the exposure is delivered at the cluster-level, care should be taken when specifying the causal model and framing the statistical estimation problem.
In particular, researchers need to consider if an individual's outcome could be impacted by another's covariates and if the cluster-level and individual $i$-specific covariates  
are sufficient to control for confounding. 
If so, the individual-level TMLE, developed under the sub-model (Figure~\ref{Figure2}), can offer asymptotic and finite sample improvements. 
If not,  estimation under the sub-model can result in misleading inference in an observational setting. 
Instead,  the cluster-level TMLE, developed under general model (Figure~\ref{Figure1}), is appropriate and can still harness the pairing of individual-level risk factors and outcomes. 
Overall, incorporating working assumptions during estimation  is more robust than assuming they hold in the underlying causal model.
For both TMLEs, the use of data-adaptive estimators, such as Super Learner, avoids the parametric modeling assumptions inherent in common multilevel approaches (e.g. random effects and GEE) and improves our chances for reliable inference.

There are several areas of future work. Examples include extensions for  missingness on the outcome vector,  longitudinal settings, and more complicated  schemes for sampling individuals within a cluster (e.g. case-control sampling). 
We plan to contrast the algorithms proposed in this manuscript with the two-stage TMLE, where an individual-level TMLE is used to obtain the optimal estimate of the cluster-level outcome $\hat{Y}^c$ (potentially accounting for informative measurement and missingness at the individual-level), and then a cluster-level TMLE  (using these cluster-level outcomes  $\hat{Y}^c$) implemented to estimate the effect of the cluster-based exposure.\citep{Balzer2012WNAR} We also plan to contrast the proposed algorithms with  augmented-GEE  \citep{Stephens2012, Stephens2014, Prague2016} when the cluster size is  informative.\citep{Seaman2014} 
Finally, we plan to generalize  the proposed algorithms to  estimate the effects of individual-level exposures in an infectious disease setting (e.g. vaccine studies).\cite{Halloran1991, Halloran1995, Morozova2018}
In all cases, the hierarchical causal models presented in this manuscript ensure that the parameter of interest is defined separately from the estimation approach and reflects the underlying scientific question. This is a distinct advantage of the Targeted Learning framework over other approaches that rely on parametric regressions to define the parameter estimated and thus the scientific question answered.\citep{MarkBook}


\section{Acknowledgements \& Funding}
The SEARCH project
gratefully acknowledges the Ministries of Health of Uganda and Kenya, our research team, collaborators
and advisory boards, and especially all communities and participants involved.
The authors also  thank Dr. Patrick Staples for his aid in the network-based simulations. We also thank the  reviewers whose comments substantially improved this manuscript.
 
 Research reported in this manuscript was supported by the National Institute of Allergy and Infectious Diseases (NIAID) of the National Institutes of Health (NIH) under 
award numbers R01AI074345, R37AI051164, and U01AI09995; and in part by  the President's Emergency Plan for AIDS Relief (PEPFAR), Bill and Melinda Gates Foundation, and Gilead
Sciences. The content is solely the responsibility of the authors and does not necessarily represent the
official views of the NIH, PEPFAR, Bill and Melinda Gates Foundation or Gilead.

\section*{Appendix A - Concrete example of the general causal model}

Consider the HIV prevention and treatment study. The general causal model (Eq.~2.1 and Figure 1 in the main text)  describes the following data generating experiment. First the unmeasured factors $U$ are drawn from $\mathbb{P}_U$. Informally, we can think of generating these background factors $U$ when we  sample the cluster from the target population and select individuals from that cluster. 
Then the community-level covariates $E$ (e.g. region, baseline HIV prevalence, perceived need) are generated by some deterministic, but unspecified, function $f_E$ of background factors $U_E$. Next the matrix of individual-level covariates $\mathbf{W}$ (e.g. demographic characteristics and risk behavior) is generated as some function $f_{\mathbf{W}}$ of the cluster-level covariates $E$ and matrix of individual-level background factors $U_{\mathbf{W}}$. This causal model specifies that the intervention $A$ may have been allocated among communities differentially and may depend on the cluster-level characteristics $E$, the matrix of individual-level characteristics ${\bf W}$, as well as the unmeasured factors included in $U_A$.  
Finally, this model assumes  that these pre-intervention community and individual-level characteristics $(E, \mathbf{W})$ together with the intervention and unmeasured factors $(A,U_\mathbf{Y})$ can affect whether each individual becomes infected with HIV by the end of the study $\mathbf{Y}$.


\section*{Appendix B -  Pooled individual-level causal effect}

When the number of sampled individuals  is constant ($N_j=n \ \forall j$), we can 
rewrite the treatment-specific mean as
\begin{equation*}
\mathbb{E} \big[ Y^{c}(a)\big] 
	= \mathbb{E}\left[ \sum_{i=1}^n \alpha_{i\LargerCdot} Y_{i\LargerCdot}(a)\right] 
	= \frac{1}{n}\sum_{i=1}^n \mathbb{E}\left[ Y_{i\LargerCdot}(a)\right]
\end{equation*}
where we have used our choice of weights $\alpha_{ij}=1/n$. 
In this case, the causal effect of the cluster-based exposure on the cluster-level outcome 
equals the average causal effect of the cluster-based exposure on the $i^{th}$ individual's outcome:
\begin{equation}\label{ATE_fixedn}
\mathbb{E}\big[ Y^{c}(1)- Y^{c}(0) \big] =\frac{1}{n}\sum_{i=1}^n  \mathbb{E}\big[ Y_{i\LargerCdot}(1) - Y_{i\LargerCdot}(0) \big]. 
\end{equation}
Further, when the index $i$ is non-informative (i.e. corresponds with  the $i^{th}$ element of a random permutation of $\{1,\ldots,n\}$), then the marginal distributions of the baseline covariates and counterfactual outcomes $({W}_{i\LargerCdot},Y_{i\LargerCdot}(1), Y_{i\LargerCdot}(0) )$ are constant in $i$. 
In this  case, the right-hand side of equation (\ref{ATE_fixedn}) does not depend on $i$ and simplifies to $\mathbb{E}\big[ Y(1) - Y(0) \big]$: the expected difference in the  individual-level counterfactual outcomes if all clusters received the treatment versus control level of the intervention.
The expectation is now over the target population of pooled individuals from all clusters. 
Applied to the HIV  example, this causal parameter (Eq.~\ref{ATE_fixedn}) evaluates the difference in the risk (probability) of HIV acquisition for a
randomly selected individual if all communities implemented the Test-and-Treat strategy versus  if all communities continued with the
standard of care.

If the number of individuals varies across clusters $(N_j \ne n \ \forall j)$, then the pooled individual-level causal effect can still be defined through an alternative cluster-level  outcome with weights as $\alpha_{ij} = J/\sum_j N_j$.
When cluster size is informative (i.e. when the intervention effect depends on the cluster size \citep{Seaman2014}),  the pooled individual-level causal effect (Eq.~\ref{ATE_fixedn}) will generally not equal the cluster-level causal effect ($\mathbb{E} \big[ Y^{c}(1) \big] -  \mathbb{E}\big[ Y^{c}(0) \big]$). Depending on the application, either or both may be of primary interest.


\section*{Appendix C - Additional details on loss functions}
 
As an initial estimator of the conditional mean outcome, we can simply regress the cluster-level outcome $Y^c$ onto the exposure and covariates 
$(A,E,{\bf W})$. We could, for example, use the squared error loss function \[
\mathcal{L}^c_{MSE}(\bar{Q}^c)(O)= \big[Y^{c}-\bar{Q}^c(A,E,{\bf W})\big]^2. \]
Alternatively, if the cluster-level outcome $Y^{c}$ is standardized so that $Y^{c}\in (0,1)$, then we could also use the binary log-likelihood loss function\citep{Gruber2010}:
\[
- \mathcal{L}^c_{ll}(\bar{Q}^c)(O)= Y^{c}\log \big[ \bar{Q}^c(A,E,{\bf W}) \big] +(1-Y^{c})\log \big[1-\bar{Q}^c(A,E,{\bf W}) \big].\]
These regressions would result in a cluster-level  analysis. For example in a linear regression model, 
the fitted regression parameters are defined as the least squares estimator:
\[
\hat{\beta} =\arg\min_{\beta}\sum_{j=1}^J \big[Y^{c}_j-\bar{Q}_{\beta}^c(A_j,E_j,{\bf W}_j)\big]^2.\]
Without making additional assumptions, 
these loss functions can also be specified at the individual-level. For the squared error loss, we have \[
\mathcal{L}_{MSE}(\bar{Q}^c)(O)=\sum_{i=1}^N \alpha_{i\LargerCdot}\big[ Y_{i\LargerCdot}-\bar{Q}^c(A,E,{\bf W})\big]^2\]
This is a valid loss function:  $\bar{Q}^c_0=\arg\min_{\bar{Q}^c}\mathbb{P}_0 \mathcal{L}_{MSE}(\bar{Q}^c)$. 
 A similar result can be proved for the binary log-likelihood loss function.
These loss functions would result in an individual-level regression analysis.  For example in a linear regression model,  the fitted regression parameters are defined as the least squares estimator:
\[
\hat{\beta}=\arg\min_{\beta} \sum_{j=1}^J \sum_{i=1}^{N_j}\alpha_{ij} \big[Y_{ij}-\bar{Q}_{\beta}^c(A_j,E_j,{\bf W}_j) \big]^2,\]
where, for example, $\alpha_{ij}=1/N_j$.
The  least squares estimator $\hat{\beta}$ solves the estimating equation:
\begin{eqnarray*}
0&=&\sum_{j=1}^J \sum_{i=1}^{N_j} \alpha_{ij} \frac{d}{d\beta} \bar{Q}_{\beta}^c(A_j,E_j,{\bf W}_j)\big(Y_{ij}-\bar{Q}_{\beta}^c(A_j,E_j,{\bf W}_j)\big)\\
&=&\sum_{j=1}^J \frac{d}{d\beta} \bar{Q}_{\beta}^c(A_j,E_j,{\bf W}_j)\left(\sum_{i=1}^{N_j} \alpha_{ij}(Y_{ij}-\bar{Q}_{\beta}^c(A_j,E_j,{\bf W}_j))\right).
\end{eqnarray*} 
From this latter equation, it follows that
the least squares estimator for the individual-level analysis is identical to the cluster-level least squares estimator. 

Under the working model assumptions (Eq.~3.7),  
the squared-error loss function for $\bar{Q}_{0}(A,E,{W}) \equiv \mathbb{E}_0(Y | A, E, {W})$ is now given by
\begin{equation*}
\mathcal{L}_{MSE}(\bar{Q})(O)=\sum_{i=1}^N \alpha_{i\LargerCdot} (Y_{i\LargerCdot}-\bar{Q}(A,E,{W}_{i\LargerCdot}) )^2.
\end{equation*}
A similar representation can be written for the  log-likelihood loss. These loss functions would result in an individual-level regression analysis, but now with paired individual-level data $(Y_{i\LargerCdot},{W}_{i\LargerCdot})$ and a much smaller adjustment set.
For example in a linear regression model, the fitted regression parameters are defined as the  least squares estimator:
\[
\hat{\beta}=\arg\min_{\beta} \sum_{j=1}^J \sum_{i=1}^{N_j}\alpha_{ij} (Y_{ij}-\bar{Q}_{\beta}(A_j,E_j,{W}_{ij} ) )^2.
\]
where, for example, $\alpha_{ij}=1/N_j$. 
Thus, we could now apply Super Learner based on this loss function  to estimate the common conditional mean function $\bar{Q}_0$, which then yields a fit of the object of interest $\bar{Q}_0^c(A,E,{\bf W})= \sum_i\alpha _{i\LargerCdot} \bar{Q}_0(A,E,{W}_{i\LargerCdot})$.
Assuming such a working model (Eq.~3.7) represents reality, an estimator of $\bar{Q}_0^c$ based on a pooled individual-level regression analysis may be more accurate than a cluster-level analysis, which is unable to pair individual-level outcomes and covariates.

\section*{Appendix D - Step-by-step implementation and \texttt{R} code}

With hierarchical data, the cluster-level TMLE for $\Psi^I(\mathbb{P}_0)$ can  be implemented in the following steps:
\begin{enumerate}

\item Estimate the expected cluster-level outcome given the exposure and covariates $\bar{Q}^c_0(A, E, \bf{W})$ using Super Learner where the library includes both cluster-level regressions and averages of individual-level regressions and where selection is based on a cluster-level loss function. 

\item Use the resulting estimator $\hat{\bar{Q}}^c$ to calculate the predicted outcomes  $\hat{\bar{Q}}^c(A_j,E_j, {\bf W}_j)$ for each cluster $j=1,\ldots,J$.
\item Estimate the cluster-level propensity score $g_0^c(a | E, \mathbf{W})$ using  parametric regression or  Super Learner with a cluster-level loss function.
\item Use the resulting estimator $\hat{g}^c$ to calculate a cluster-level clever covariate $\hat{H}_j^c=\frac{\mathbb{I}(A_j=a)}{\hat{g}^c(A_j \mid E_j,{\bf W}_j)}$ for each cluster $j=1,\ldots,J$.
\item Estimate the fluctuation coefficient  $\epsilon$ by running parametric logistic regression of the cluster-level outcome $Y^{c}$ on the cluster-level  covariate $\hat{H}^c$ with offset as $logit(\hat{\bar{Q}}^{c})$.
\item Obtain targeted  predictions of the cluster-level outcome 
as  \[
\hat{\bar{Q}}^{c*}(a,E_j,{\bf W}_j)=logit^{-1}\big[ logit[\hat{\bar{Q}}^c(a,E_j, {\bf W}_j)]+\hat{\epsilon}  \hat{H}_j^c\big] \]
 for each cluster $j=1,\ldots,J$. 
 \item Obtain a point estimate  by taking the empirical mean of these targeted predictions across the sample of $J$ clusters: \[
\hat{\Psi}^I(Q^*)(a) = \frac{1}{J}\sum_{j=1}^J \hat{\bar{Q}}^{c*}(a,E_j,{\bf W}_j).\]
\item Construct 95\% confidence intervals for the resulting TMLE as $\hat{\Psi}^I  \pm 1.96 \times \frac{\hat{\sigma}}{\sqrt{J}}$ where $\hat{\sigma}^2$ is the sample variance of the estimated influence curve $\hat{D}^{I}(\hat{Q}^*,\hat{g}^c)$ (Eq.~3.6 in main text).\\
 \end{enumerate}

\noindent The individual-level TMLE for $\Psi^{II}(\mathbb{P}_0)(a)$ can be implemented in the following steps: 

\begin{enumerate}

\item Estimate the expected individual-level outcome given the exposure and covariates $\bar{Q}_0(A, E, {W})$ using Super Learner where the library includes parametric and data-adaptive pooled individual-level regressions and where selection is based on a individual-level loss function. 
 If cluster size varies, include weights $\alpha_{ij}=1/N_j$.
 
\item Use the resulting estimator $\hat{\bar{Q}}$ to calculate the predicted outcomes $\hat{\bar{Q}}(A_j, E_j, {W}_{ij})$ for each individual $i=1,\ldots,N_j$ in cluster $j=1, \ldots, J$. 
 
\item Estimate the individual-level propensity score $g_0(a | E, {W}_{i\LargerCdot})$  using a pooled individual-level regression of $A$ on $(E, {W}_{i\LargerCdot})$ or using more data-adaptive methods, such as Super Learner, with a  individual-level loss function. If cluster size varies, include weights $\alpha_{ij}=1/N_j$.

\item Use the resulting estimator $\hat{g}$ to calculate an individual-level clever covariate $\hat{H}_{ij}=\frac{\mathbb{I}(A_j=a)}{\hat{g}(A_j \mid E_j, {W}_{ij})}$ for each individual $i=1,\ldots,N_j$ in cluster $j=1,\ldots,J$.

\item Estimate the fluctuation coefficient  $\epsilon$ by running pooled parametric logistic regression of the individual-level outcome $Y_{i\LargerCdot}$ on the individual-level  covariate $\hat{H}_{i\LargerCdot}$ with offset as $logit(\hat{\bar{Q}})$. If cluster size varies, include weights $\alpha_{ij}=1/N_j$.

\item Use the targeted estimator to obtain  predictions of the individual-level outcome $Y_{i\LargerCdot}$ given $A=a$ and covariates as  \[
\hat{\bar{Q}}^*(a,E_{j}, W_{ij})=logit^{-1}\big[ logit[\hat{\bar{Q}}(a,E_{j},  W_{ij})]+\hat{\epsilon}  \hat{H}_{ij}\big] \]
 for each individual $i$ in each cluster $j$. 
 
 \item Obtain a point estimate  by taking the empirical mean of these targeted predictions within clusters and then across the sample of $J$ clusters: \[
\hat{\Psi}^{II}(\hat{Q}^*)(a)= \frac{1}{J}\sum_{j=1}^J  \sum_{i=1}^{N_j} \alpha_{ij} \hat{\bar{Q}}^*(a,E_{j}, W_{ij}).
\]
\item  Construct 95\% confidence intervals for the resulting TMLE as $\hat{\Psi}^{II} \pm 1.96 \times \frac{\hat{\sigma}}{\sqrt{J}}$ where $\hat{\sigma}^2$ is the sample variance of the estimated influence curve $D^{II}(\hat{Q}^*,\hat{g})$. \\
 \end{enumerate}

\noindent Full \texttt{R} code for the simulations and estimators is at \url{https://github.com/LauraBalzer/HierarchicalTMLE}.

\section*{Appendix E- Theoretical comparison of the TMLEs}

\begin{proof}
 Suppose that the true observed data distribution $\mathbb{P}_0$ is an element of the sub-model $\mathcal{M}^{II}$. Then we have $\Psi^I(\mathbb{P}_0)(a) = \Psi^{II}(\mathbb{P}_0)(a) = \psi_0(a)$. 
For simplicity, also consider a randomized trial with $g_0^c(A | E, \mathbf{W})=g_0(A|E, W) = 0.5$. Then we can re-write the efficient influence curves as  
\begin{equation}
D^{I}(\mathbb{P}_0)(O) = 2\mathbb{I}(A=a) \big( Y^{c}-\bar{Q}_{0}^c(A,E,{\bf W}) \big) +\bar{Q}_{0}^c(a,E,{\bf W})-\psi_0(a)
\end{equation}
and 
\begin{equation}
D^{II}(\mathbb{P}_0)(O)=  \sum_{i=1}^N  \left[ \alpha_{i\LargerCdot} 2 \mathbb{I}(A=a) 
\big(Y_{i\LargerCdot}-\bar{Q}_0(A,E,{W}_{i\LargerCdot})\big) + \bar{Q}_0(a,E,{W}_{i\LargerCdot})-\psi_0(a) \right]
\end{equation}
Due to the linearity of summations, one can show that in this setting $D^{I}(\mathbb{P}_0)(O) =D^{II}(\mathbb{P}_0)(O)$ and thus the efficiency bound is the same.
\end{proof}

\FloatBarrier

\section*{Supplementary figures}

\begin{figure} 
\centering
\includegraphics[width=.3\textwidth]{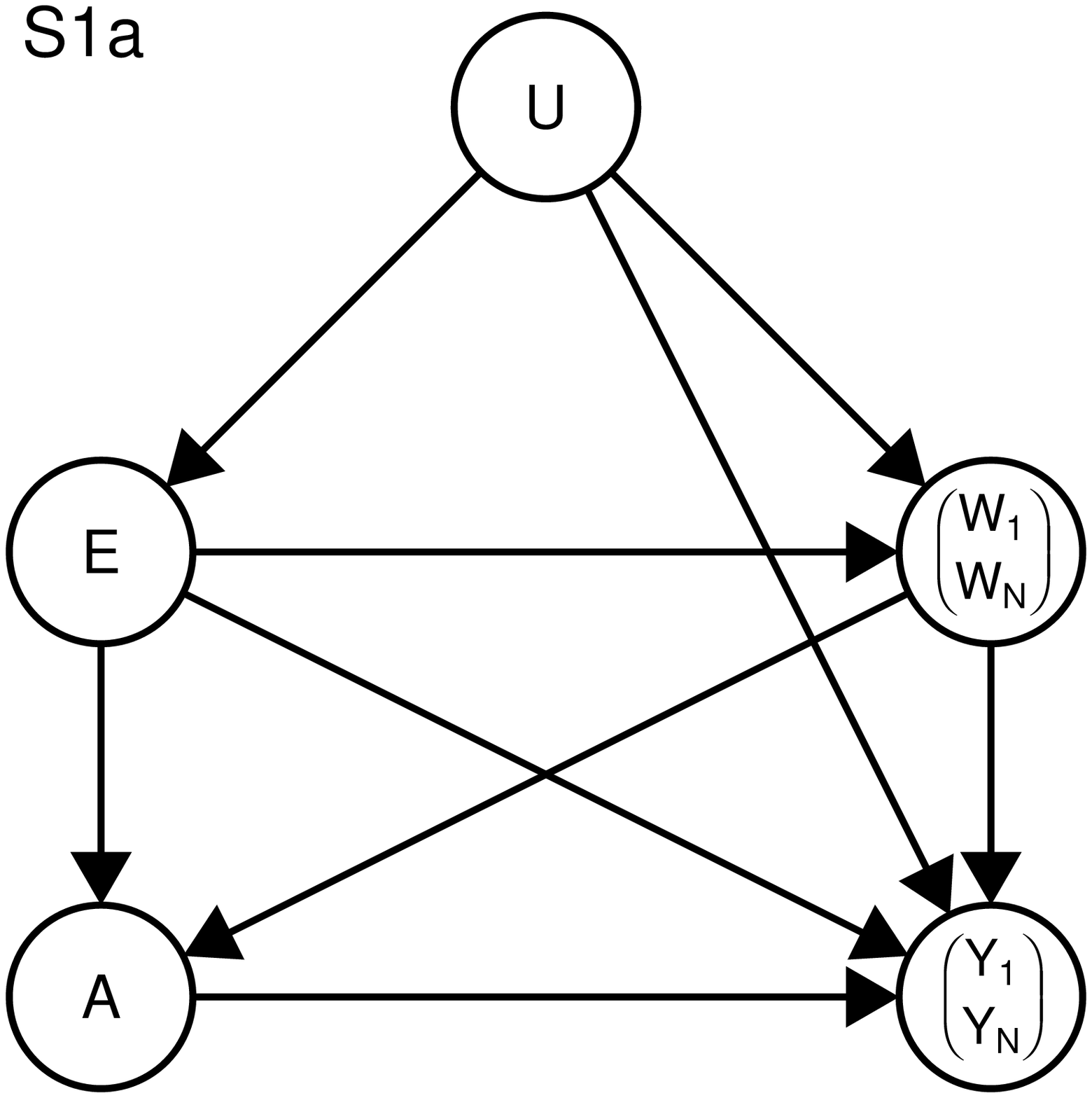} 
\hspace{1em}
\includegraphics[width=.3\textwidth]{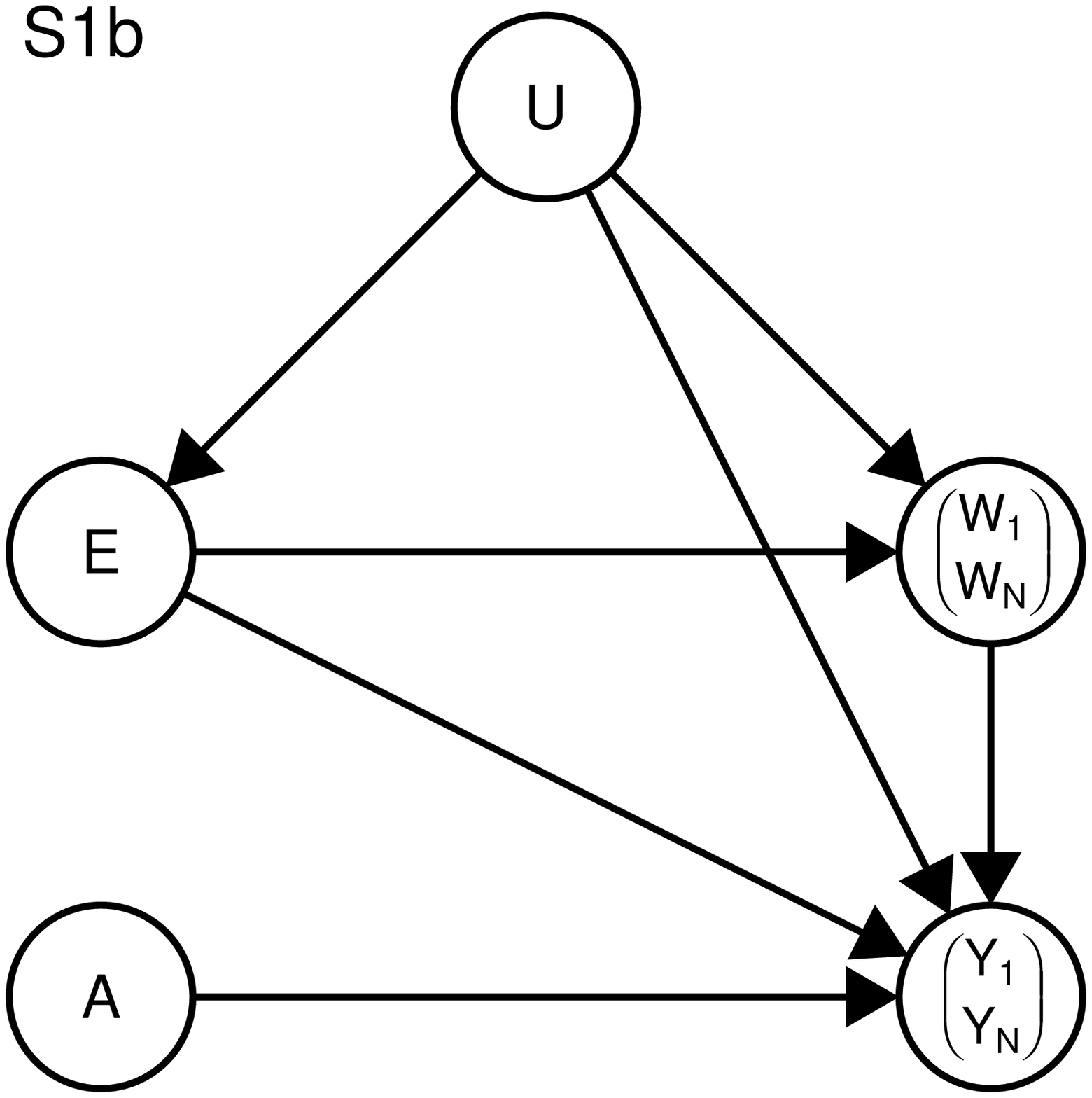}
\caption{Two possible directed acyclic graphs (DAGs) that are compatible with the no unmeasured confounders assumption in the general causal model. Here,  $U$ denotes unmeasured factors, $E$ the cluster-level covariates, $(W_{1.},\ldots W_{N.})$ the individual-level covariates, $A$  the cluster-level exposure, and $(Y_{1.},\ldots, Y_{N.})$ the individual-level outcomes. \textbf{S1a:} an observational setting where the covariates ($E, (W_{1.},\ldots W_{N.})$) are sufficient to control for confounding. \textbf{S1b:} cluster randomized trial where by design there is no confounding.}
\label{FigureA}
\end{figure}

\begin{figure} 
\centering
\includegraphics[width=.3\textwidth]{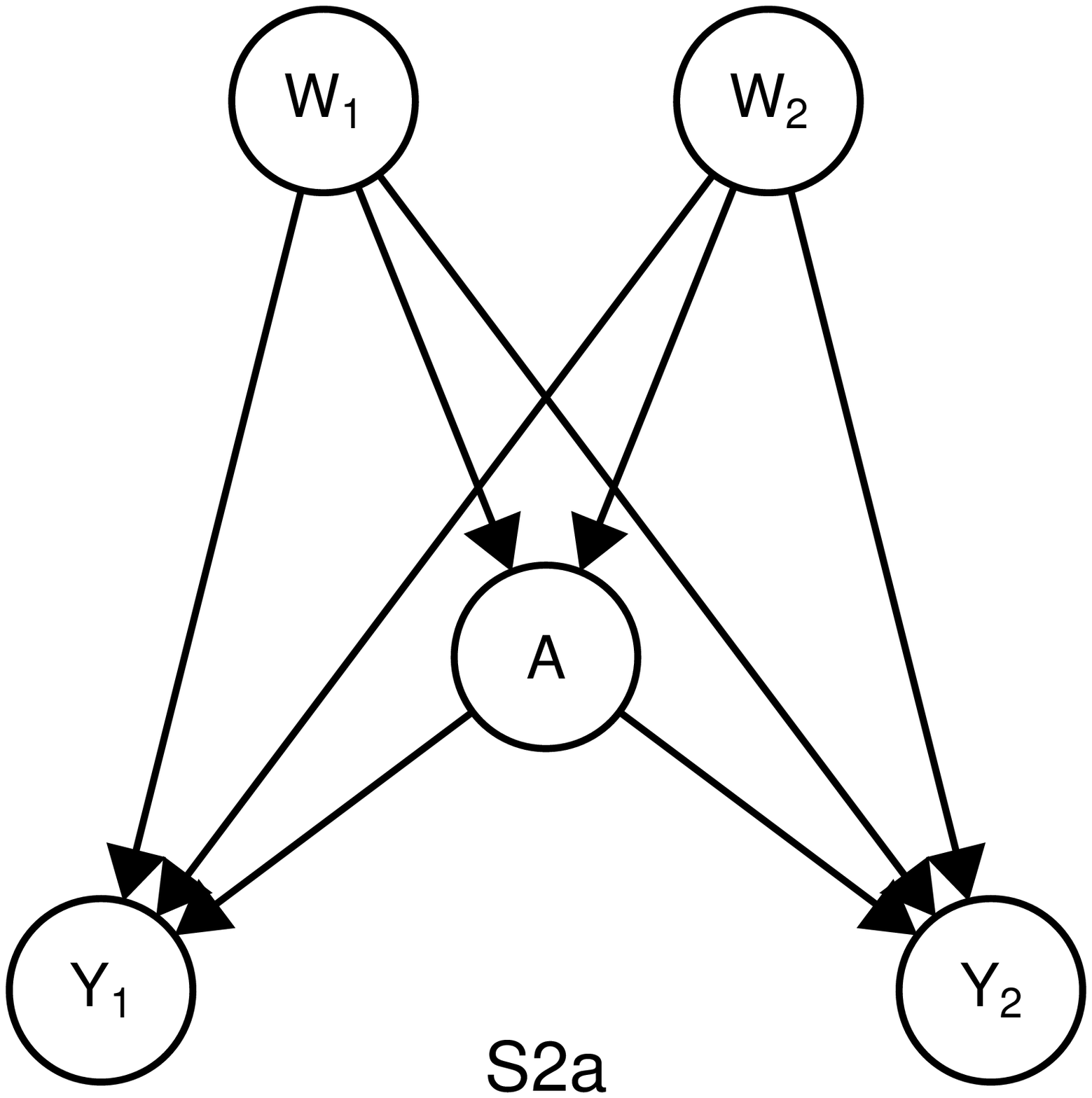} 
\hspace{.25em}
\includegraphics[width=.3\textwidth]{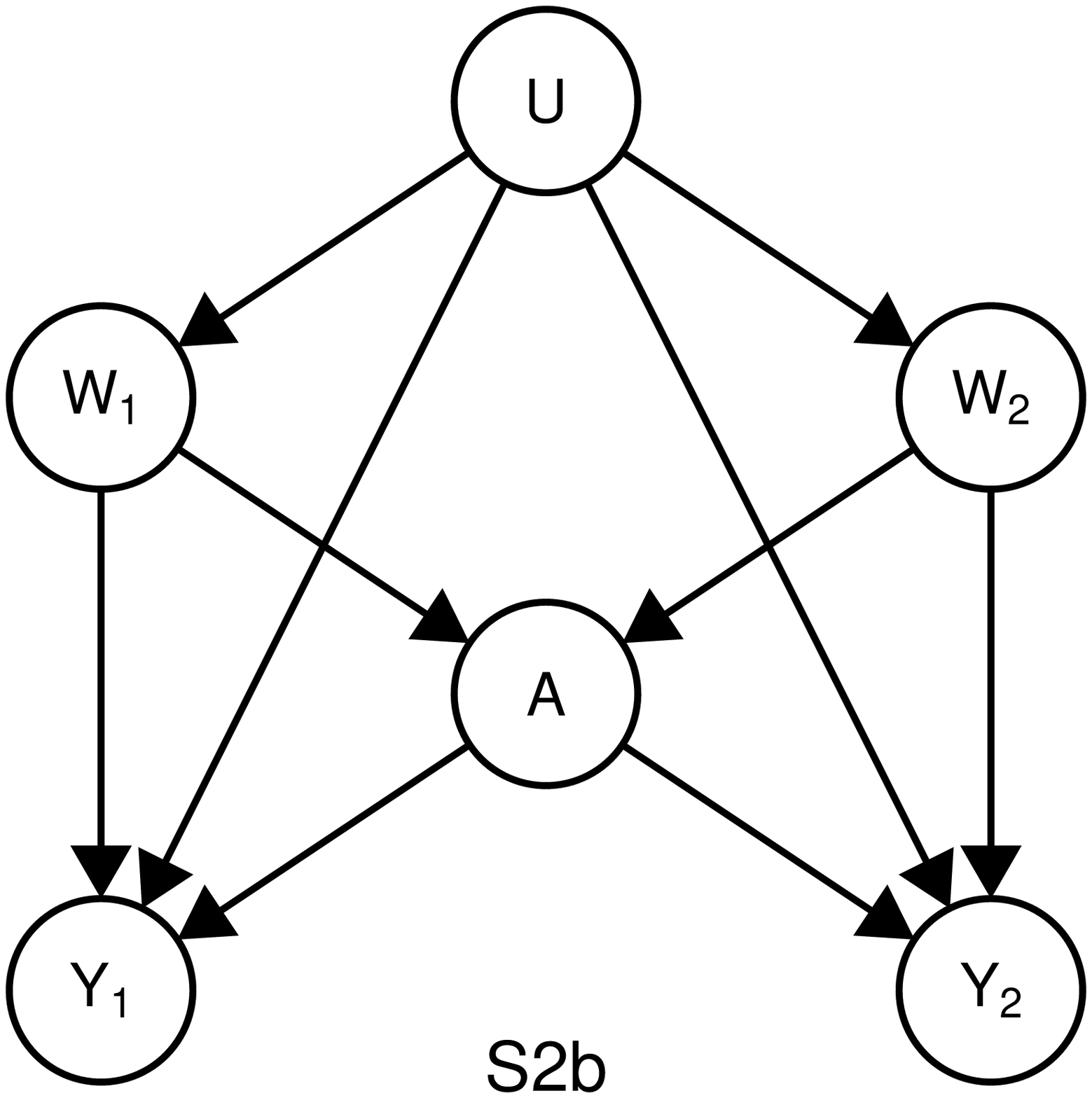} 
\hspace{.25em}
\includegraphics[width=.3\textwidth]{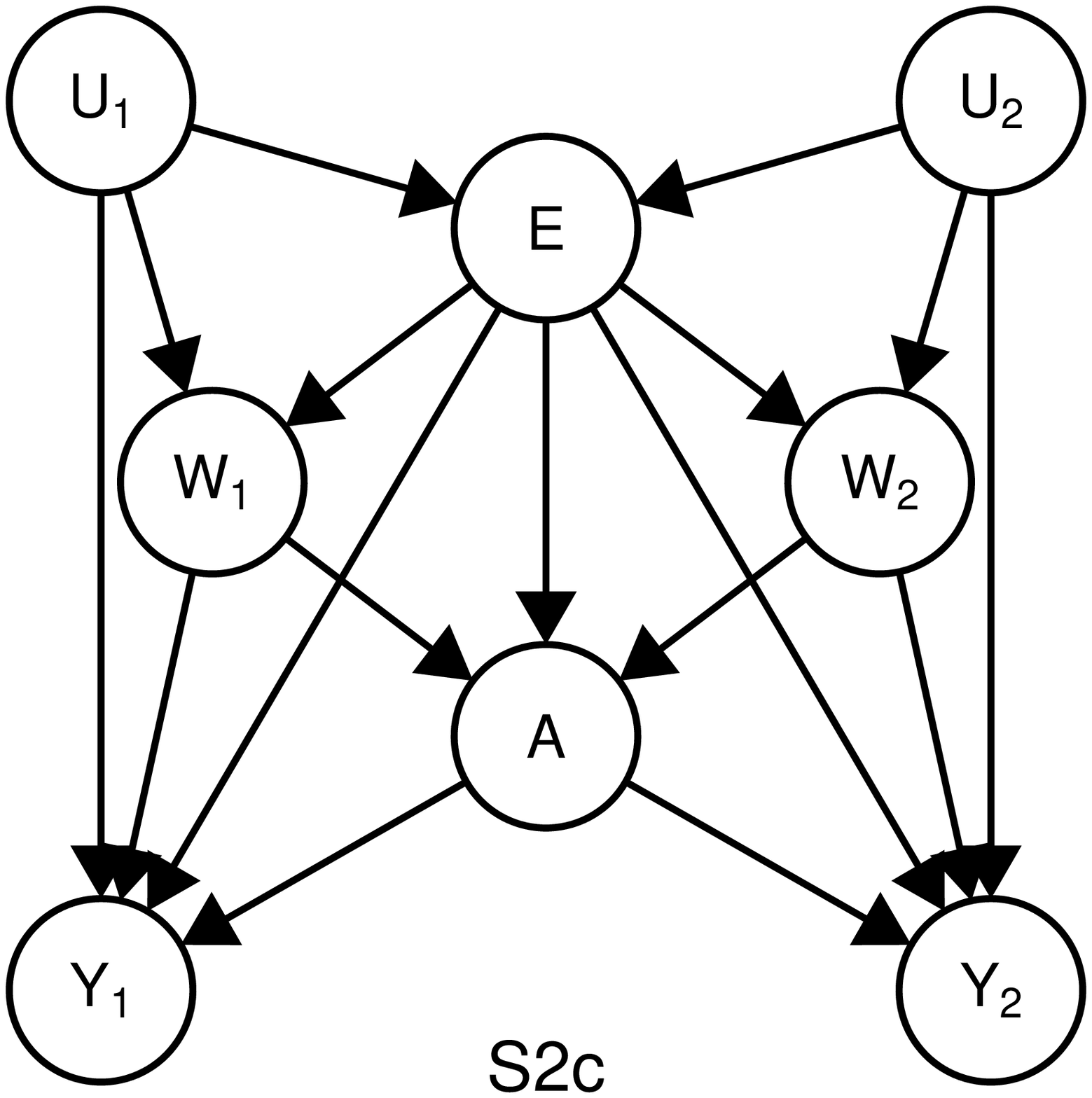}
\caption{Directed acyclic graphs (DAGs) to illustrate the assumptions on the distribution of unmeasured factors. Let $U$ denote unmeasured factors, $E$ the cluster-level covariates, $W$ the individual-level covariates, $A$  the cluster-level exposure, and $Y$ the individual-level outcome. For ease of presentation, we only show two individuals, denoted by subscripts 1 and 2, in a given cluster. In all causal models, the measured covariates capture all the common causes of the exposure and outcomes. 
\textbf{S2a:}  For simplicity, we ignore the cluster-level covariates $E$. Even if all the unmeasured factors are independent (and thus not explicitly shown), we need to control for both $(W_{1.}, W_{2.})$ when there is  covariate interference (i.e $Y_{1.}$ is  a function of $W_{2.}$ and $Y_{2.}$ is  a function of $W_{1.}$). The assumptions in the restricted causal model do not hold. 
\textbf{S2b:} For simplicity, we again ignore the cluster-level covariates $E$. Even with no covariate interference, we need to control for both $(W_{1.}, W_{2.})$ when there is a shared unmeasured common cause of the individual-level covariates and individual-level outcomes.  The assumptions in the restricted causal model do not hold. 
 \textbf{S2c:} Let $U_{1.}$ and $U_{2.}$ denote the $i$-specific unmeasured common causes of the cluster-level covariates, individual-level covariates, and individual-level outcome.  Even with no covariate interference, we need to control for $(E, W_{1.}, W_{2.})$, because the cluster-level covariates $E$ are  a collider of the $U_{1.}$ and $U_{2.}$. The assumptions in the restricted causal model do not hold. 
 }
\label{FigureB}
\end{figure}

\begin{figure} 
\centering
\includegraphics[width=.3\textwidth]{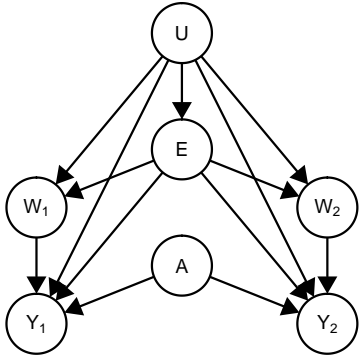}
\caption{ When the cluster-level exposure is randomized,  we do not need to adjust for covariates, regardless of the error structure. If there is also no covariate interference, the assumptions in the restricted causal model do  hold. 
}
\label{FigureC}
\end{figure}

\FloatBarrier

\section*{Supplementary tables}

\begin{table}[!htbp]
\small\sf\centering
\caption*{Supplementary Table S1: True value of the causal effect of the cluster-level exposure $\mathbb{E}[ Y^c(1) - Y^c(0)]$ for each of the data generating processes in Simulation 1. When there is a treatment effect, the coefficient for the exposure in the logistic regression for the conditional probability of the \textbf{individual-level} outcome (Eq. 6.1-6.2) is 0.1. Nonetheless, the strength of the effect of the cluster-level exposure on the \textbf{cluster-level} outcome depends on the presence or absence of strong covariate interference as well as the presence or absence of dependence in the unmeasured factors determining the individual-level outcomes $U_{\mathbf{Y}}$. By construction, the treatment effect is always 0 in the null setting.  All measures are in \%. }
\begin{tabular}{l ll ll}
  \toprule
              & \multicolumn{2}{c}{With an effect}  &  \multicolumn{2}{c}{Under the Null}  \\
				& Indpt.~$U_\mathbf{Y}$ & Dept.~$U_\mathbf{Y}$  & Indpt.~$U_\mathbf{Y}$ & Dept.~$U_\mathbf{Y}$  \\
Minimal covariate interference &  1.6 & 3.8 & 0 & 0 \\
Stronger covariate interference &  2.1 & 6.3 & 0 & 0 \\
\bottomrule
\end{tabular}
\label{Table:Sim1Parametr}
\end{table}

\begin{table}[!htbp]
\small\sf\centering
\caption*{Supplementary Table S2: Estimator performance in Simulation 1 under minimal covariate interference (Eq.~6.1)  and under stronger covariate interference  (Eq.~6.2). We also vary the dependence of the unmeasured factors determining the individual-level outcomes: independent (top) and correlated (bottom). 
Performance is given by
bias as the average deviation between the estimate and truth;  $\sigma$ as the standard error; rMSE as the root-mean squared error; 
 type I error as the proportion of times the true null hypothesis is rejected, and coverage as the proportion of times the 95\% confidence interval contains the true value.
 All measures are in \%. }
\begin{tabular}{l lllll lllll }
  \toprule
              & \multicolumn{5}{l}{Minimal covariate interference}  &  \multicolumn{5}{l}{Stronger covariate interference }  \\
Estimator &   Bias & $\sigma$ & rMSE & Type I & Coverage &    Bias & $\hat{\sigma}$ & rMSE & Type I & Coverage \\
\midrule
Unadj. & 10.4 & 5.1 & 11.5 & 54 & 46 & 7.6 & 3.9 & 8.5 & 51 & 49 \\ 
  TMLE-$Ia$ & -0.0 & 1.2 & 1.2 & 6 & 94 & -0.0 & 1.4 & 1.4 & 6 & 94 \\ 
  TMLE-$Ib$ & -0.0 & 1.2 & 1.2 & 5 & 95 & -0.0 & 1.4 & 1.4 & 2 & 98 \\ 
  TMLE-$II$ & 0.2 & 1.2 & 1.2 & 6 & 94 & 1.6 & 1.6 & 2.3 & 18 & 82 \\ 
    & \multicolumn{8}{c}{Independent $U_{\mathbf{Y}}$ determining the outcome } \\
\midrule
Unadj. & 6.5 & 3.3 & 7.3 & 53 & 47 & -3.8 & 2.5 & 4.5 & 34 & 66 \\ 
  TMLE-$Ia$ & -0.0 & 1.3 & 1.3 & 5 & 95 & 0.0 & 1.8 & 1.8 & 6 & 94 \\ 
  TMLE-$Ib$ & -0.0 & 1.3 & 1.3 & 0 & 100 & 0.0 & 1.8 & 1.8 & 2 & 98 \\ 
  TMLE-$II$ & -4.2 & 2.3 & 4.8 & 43 & 57 & -2.3 & 2.1 & 3.1 & 19 & 81 \\ 
    & \multicolumn{8}{c}{Dependent $U_{\mathbf{Y}}$ determining the outcome } \\
\bottomrule
\end{tabular}
\label{Table:Sim1ResultsNull}
\end{table}

\begin{table}[!htbp]
\small\sf\centering
\caption*{Supplementary Table S3: For the TMLEs developed under the general model $\mathcal{M}^I$ and under the sub-model $\mathcal{M}^{II}$, the number of times a candidate variable was selected for adjustment during initial estimation of the outcome regression or the known propensity score in Simulation 2. The candidates include nothing (``Unadj."), degree, demographic risk group (``Demo."), the  number of partners infected at baseline (``N. partners"), cluster-level baseline HIV prevalence, assortativity (``Assort."), and the number of distinct sexual groups (``N. components").
}
\begin{tabular}{l ccccccc}
\toprule
& Unadj. & Degree & Demo. & N. partners & Prevalence & Assort. & N. components \\ 
  \midrule
   \multicolumn{8}{l}{\textbf{Selection under the general model (TMLE-$\mathcal{M}^I$)}}\\
 Outcome regression  & 2 & 64 & 4 & 759 & 112 & 8 & 51 \\ 
  Propensity& 830 & 36 & 38 & 8 & 33 & 25 & 30 \\ 
  \multicolumn{8}{l}{\textbf{Selection under the sub-model (TMLE-$\mathcal{M}^{II}$)}} \\
Outcome regression  & 2 & 64 & 4 & 759 & 112 & 8 & 51 \\
Propensity score & 877 & 14 & 6 & 8 & 33 & 26 & 36 \\ 
 \bottomrule
\end{tabular}
\label{Table:Sim2Select}
\end{table}

\FloatBarrier

\bibliography{/Users/laurabalzer/Dropbox/Research/Bibliography}

\end{document}